\documentclass[aps,prb,reprint,noeprint,superscriptaddress]{revtex4-2}
\usepackage[utf8]{inputenc}
\usepackage[english]{babel}
\usepackage{microtype}
\usepackage{amsmath,amssymb,amsfonts}
\usepackage{braket}
\usepackage{bm}
\usepackage{graphicx}
\usepackage{booktabs}
\usepackage{comment}
\usepackage{siunitx}
\usepackage{float}
\usepackage[
    pdftitle={},
    pdfauthor={},
    colorlinks=true,
    unicode=true,
    pdfborder={0 0 0},
    allcolors=blue
]{hyperref}

\renewcommand{\vec}[1]{\bm{#1}}

\renewcommand{\Re}{\operatorname{\mathsf{Re}}}
\newcommand{\vk}{{\vec k}}

\newcommand{\VV}[2]{\begin{pmatrix}#1\\#2\end{pmatrix}}
\newcommand{\VVT}[2]{\begin{pmatrix}#1 & #2\end{pmatrix}}

\newcommand{\MM}[4]{\begin{pmatrix}#1 & #2 \\ #3 & #4\end{pmatrix}}

\begin{document}

\title{Time-resolved optical conductivity and Higgs oscillations\\in two-band dirty superconductors}

\author{Rafael Haenel}
\affiliation{Max Planck Institute for Solid State Research,
70569 Stuttgart, Germany}
\affiliation{Department of Physics and Astronomy, University of British Columbia, Vancouver V6T 1Z4, Canada}
\affiliation{Quantum Matter Institute, University of British Columbia, Vancouver V6T 1Z4, Canada}

\author{Paul Froese}
\affiliation{Max Planck Institute for Solid State Research,
70569 Stuttgart, Germany}
\affiliation{Department of Physics and Astronomy, University of British Columbia, Vancouver V6T 1Z4, Canada}

\author{Dirk Manske}
\affiliation{Max Planck Institute for Solid State Research,
70569 Stuttgart, Germany}

\author{Lukas Schwarz}
\affiliation{Max Planck Institute for Solid State Research,
70569 Stuttgart, Germany}

\date{\today}

\begin{abstract}
Recent studies have emphasized the importance of impurity scattering for the
optical Higgs response of superconductors. In the dirty limit, an additional
paramagnetic coupling of light to superconducting condensate arises which
drastically enhances excitation. So far, most work concentrated on the periodic
driving with light, where the third-harmonic generation response of the Higgs
mode was shown to be enhanced. In this work, we additionally calculate the time-resolved
optical conductivity of single- and two-band superconductors in a two-pulse
quench-probe setup, where we find good agreement with existing experimental
results. We use the Mattis-Bardeen approach to incorporate impurity scattering
and calculate explicitly the time-evolution of the system. Calculations are performed
both in a diagrammatic picture derived from an effective action
formalism and within a time-dependent density matrix formalism.
\end{abstract}

\maketitle

\section{Introduction}
\label{sec:introduction}

When a continuous symmetry is spontaneously broken, collective excitations
emerge. In the case of a superconductor, where the complex order parameter
$\Delta e^{i\varphi}$ acquires a finite value below a critical temperature
$T_C$, two bosonic modes appear: the massive Higgs mode and a massless
Goldstone mode \cite{Varma2002,Pekker2015}. They may be seen as
amplitude $\delta\Delta$ and phase $\delta\theta$ fluctuations of the
complex order parameter in the Mexican hat-shaped free energy potential. In the
presence of Coulomb interaction, the Goldstone mode is shifted to optical
frequencies by means of the Anderson-Higgs mechanism while the Higgs mode
remains a stable gapped excitation in the Terahertz regime \cite{Anderson1958}.

In a two-band superconductor, two gapped Higgs modes and two phase modes exist.
While the global phase fluctuation occurs again only at energies close to the
plasma frequency in the presence of long-range Coulomb interactions, the
relative phase fluctuation, quantized as the Leggett mode, persists as a gapped
excitation at low energies \cite{Leggett1966}.

Experimental observation of Higgs and Leggett collective modes is difficult.
Since these fields are scalar quantities, no linear coupling to the
electromagnetic field exists \cite{Pekker2015}. Thus, there are no direct
experimental signatures in linear response. Only with a coexisting CDW order, a
signature of the Higgs mode is visible in Raman spectroscopy
\cite{Sooryakumar1980,Littlewood1981,Littlewood1982,Measson2014,Cea2014}.
As a consequence, experiments need to be performed in the non-linear regime. Here, the
challenge is twofold: intense light sources are required but experiments also
have to be performed on energy scales mostly within the superconducting gap
such that optical excitation of quasiparticles does not deplete the condensate.

Recent developments in ultrafast Terahertz spectroscopy have caused a surge in
interest to study collective excitations in non-equilibrium superconductors both
in theory
\cite{Papenkort2007,Krull2014,Tsuji2015,Krull2016,Kumar2019,Schwarz2020} and
experiment
\cite{Matsunaga2013,Matsunaga2014,Katsumi2018,Giorgianni2019,Chu2020,Katsumi2020,Kovalev2020,Vaswani2020},
where first experimental signatures of the Higgs mode have been observed in
various materials.
The main excitation scheme so far consist of two approaches. First, samples are
illuminated in a pump-probe setup where an excitation of the Higgs mode by a
single-cycle THz pump acting as a quench appears as an oscillation of the probe
signal as a function of pump-probe delay \cite{Matsunaga2013}. In a second type
of experiments, the Higgs mode is resonantly driven by an intense multi-cycle
pulse that yields an electrical field component of three times the pump
frequency in the reflected or transmitted beam
\cite{Matsunaga2014,Chu2020,Kovalev2020}.

The fact that characteristics of the Higgs mode in superconductors are
observable in experiments is not self-evident. Early
theoretical calculations in the clean limit predicted extremely weak
experimental signatures that relied on breaking of the particle-hole symmetry.
Therefore, the first observations \cite{Matsunaga2014} of the third-harmonic
response generated by the Higgs mode was doubted \cite{Cea2016} as it should be
overlaid by much stronger charge fluctuations. Only recently, the role of
impurities has been appreciated as it drastically enhances
the coupling of light to the Higgs mode due to an additional paramagnetic
coupling absent in the clean limit
\cite{Murotani2019,Silaev2019,Tsuji2020,Seibold2020}.
This coupling becomes the dominant contribution even for small disorder.
It was further shown that impurity scattering yields qualitatively different behavior
in the polarization dependence of the driving pulses \cite{Seibold2020}.

While previous studies on impurities concentrated mostly on the excitation
scheme with periodic driving, in this work, we additionally explore the excitation with a
two-pulse quench-probe scheme. We consider both one- and two-band
superconductors where the bands can be in different impurity regimes.

We also calculate the individual contributions of quasiparticles, Leggett mode
and Higgs mode to the third-harmonic generation response. Our results support
the findings of a recent work, where the third-harmonic response in the
two-band superconductor MgB$_2$ shows a resonance only for the lower gap
\cite{Kovalev2020}. This can be understood from the fact that the upper band is
either in the clean limit or that the Fermi surface is very small.

We incorporate the effect of impurities in our model using the Mattis-Bardeen
approximation \cite{Mattis1958}. This approach constitutes an excellent description for many 
conventional superconductors \cite{Seibold2007mb}. To
calculate the time-resolved optical conductivity, we extend the density-matrix
approach of \cite{Murotani2019} to a two-pulse excitation scheme. Here, the
short first pulse acts as a quench, while the second probe pulse with variable
time-delay probes the dynamics of the system. In addition, we consider
a diagrammatic approach derived from an effective action formalism, where the
Mattis-Bardeen ansatz is incorporated by an effective finite momentum
interaction vertex. This diagrammatic approach is equivalent to the
density matrix formalism but allows to understand the involved processes in
more detail.

This article is organized as follows. In Sec.~\ref{sec:model} we formulate the
model (a) in terms of a diagrammatic expansion of an effective action and (b) in terms of a
density matrix equation of motion approach that was previously established by
Murotani and Shimano \cite{Murotani2019}. The two formulations are equivalent. 
We then discuss results in the case of a single-band superconductor
in a pump-probe scenario in Sec.~\ref{sec:singleband}. In Sec.
\ref{sec:multiband} we study in detail the case of a two-band superconductor
motivated by material parameters of MgB$_2$. Here we focus on both pump-probe
and third harmonic generation (THG) experiments. We summarize all results in
Sec.~\ref{sec:conclusion}.

\section{Model}
\label{sec:model}
\subsection{Hamiltonian}
We consider the BCS multiband Hamiltonian
\begin{align}
    \mathcal{H}_0 &= \sum_{i\mathbf{k}\sigma}^{}\epsilon_{i\mathbf{k}}
    c_{i\mathbf{k}\sigma}^\dagger c_{i\mathbf{k}\sigma}
     - \sum_{ij\mathbf{kk'}}^{}U_{ij} c_{i\mathbf{k}\uparrow
    }^\dagger c_{i-\mathbf{k}\downarrow }^\dagger c_{j-\mathbf{k'}\downarrow
    }c_{j\mathbf{k}'\uparrow }
\end{align}
where $\epsilon_{i\mathbf{k}} = s_i \left(\mathbf{k}^2/2m_i -
\epsilon_{F_i}\right)$ is the parabolic dispersion of the $i$-th band with
Fermi-energy $\epsilon_{F_i}$ and electron mass $m_i$. The factor $s_i=\pm$
determines electron- or hole-like character of the respective band.

At the mean-field level the interacting term is decoupled in the pairing channel,
\begin{align}
  \sum_{i\mathbf{k}}^{}
    \Delta_i c_{i\mathbf{-k}\uparrow }^\dagger
    c_{i\mathbf{k}\downarrow }^\dagger   +\textit{h.c.} \,,
\end{align}
where order parameters $\Delta_i$ are self-consistently determined by the
BCS gap equation $\Delta_i = \sum_{j\mathbf{k}}^{}U_{ij} \langle
c_{j-\mathbf{k}\downarrow }c_{j\mathbf{k}\uparrow }\rangle$ \cite{Suhl1959}.
The order parameters of different bands are mixed by off-diagonal terms in the
coupling matrix $U_{ij}$. In the present work, we parametrize gap-mixing by a
parameter $v$ and define
\begin{align}
  U_{ij} =
  \begin{pmatrix}
    U_{11} & v U_{11} \\
    v U_{11} & U_{22}
  \end{pmatrix} \,.
  \label{eqn:v}
\end{align}
For given $\Delta_i$ and $v$ we can find $U_{11}$ and
$U_{22}$ such that the gap equation is satisfied.

To model an experimental probe with a laser pulse, we introduce a
time-dependent vector potential $\mathbf{A}(t)=A(t)\, \mathbf{e}$ with
polarization vector $\mathbf e$ by means of minimal coupling,
\begin{align}
    \mathcal{H}_{1} = -\sum_{i\mathbf{kk'}\sigma}^{}
    \mathbf{J}_{i\mathbf{kk'}} \cdot \mathbf{A} \,
    c_{i\mathbf{k}\sigma}^\dagger  c_{i\mathbf{k}'\sigma} +
    \sum_{i\mathbf{k}\sigma}^{} \frac{s_i e^2}{2m_i} \mathbf{A}^2 \,
    c_{i\mathbf{k}\sigma}^\dagger c_{i\mathbf{k}\sigma}\,,
\end{align}
where $J_{i\mathbf{kk'}}=\bra{i\mathbf{k}}\frac{e\mathbf{p}_i}{m_i}
\ket{i\mathbf{k'}}$ are intraband transition matrix elements of the current
operator. The two terms in $\mathcal{H}_1$ corresponds to the paramagnetic and
diamagnetc coupling of the laser field, respectively. The full Hamiltonian is
given by $\mathcal{H} = \mathcal{H}_{0} + \mathcal{H}_{1}$.

\subsection{Impurity scattering}
\label{sec:MB}
In a clean system momentum conservation yields $\mathbf{J}_{i\mathbf{kk'}} \sim
\delta_{\mathbf{kk'}}$, or $\mathbf{J}_{i\mathbf{kk'}} \sim
\delta_{\mathbf{k,k'}\pm\mathbf{q}}$ if a photon wavevector $\mathbf{q}$ is
considered. In disordered systems, translational invariance is broken, so that
transitions between states of different momenta are allowed.
Here, we adopt the approach of Murotani and Shimano \cite{Murotani2019} and
model the effects of impurities within the Mattis-Bardeen (MB) approximation
\cite{Mattis1958}. Explicitly, impurities enter through the approximation
\begin{align}
    \langle \left|\mathbf{e} \cdot
    \mathbf{J}_{i\mathbf{kk'}}\right|^2\rangle_{\text{Av}}
    &= \int \frac{d\Omega_\mathbf{k}}{4\pi} \frac{d\Omega_{\mathbf{k}'}}{4\pi}
    \left|\mathbf{e} \cdot \mathbf{J}_{i\mathbf{kk'}}\right|^2
    \nonumber
    \\
    &\approx \frac{(e v_{F_i})^2}{3 N_i(0)}
    W(\epsilon_{i\mathbf{k}},\epsilon_{i\mathbf{k}'})
     \,,
        \label{eq:MB}
     \\
     W(\epsilon_{i\mathbf{k}},\epsilon_{i\mathbf{k}'})
     &= \frac{1}{\pi}\frac{\gamma_i}{\left(\epsilon_{i|\mathbf{k}|}-\epsilon_{i|\mathbf{k}'|}\right)^2
    + \gamma_i^2}
    \label{eq:W-lorentz}
\end{align}
with Fermi velocity $v_{F_i}$, density of states at the Fermi level $N_i(0)$
and impurity scattering rate $\gamma_i$. A derivation of this matrix element is
given in Ref.~\cite{Murotani2019}.

We see that impurity scattering broadens the
$\delta_{\mathbf{kk'}}$-distribution into a Lorentzian of width $\gamma_i$
centered at zero momentum transfer. The bandstructure defined by
$\mathcal{H}_0$ remains unaffected in this approximation. Instead of broadening
the momentum resolution of the bandstructure, one may view impurities as
effectively broadening the momentum of the photon.

\subsection{Effective Action}
\begin{figure}[ht]
    \centering
    \includegraphics[width=0.9\columnwidth]{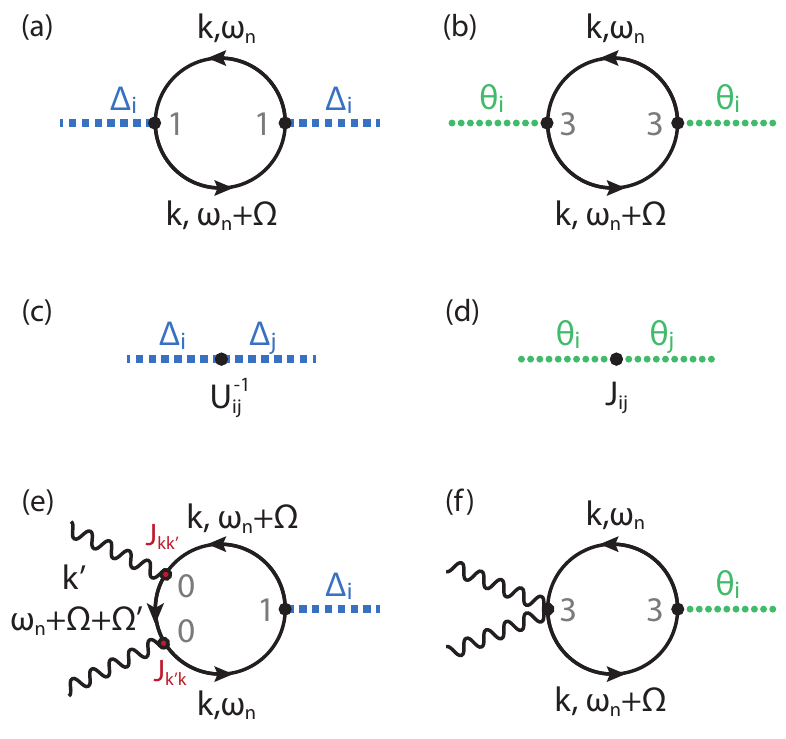}
    \caption{Diagrammatic representation of terms in the quadratic effective action $S[\Delta_i,\theta_i, \vec A]$ in Eq.~\eqref{eq:seffappdx} involving Higgs fields (left column) and phase fields (right column). 
    Bubbles correspond to susceptibilities listed in Eqs.~\eqref{eq:susc-begin}-\eqref{eq:susc-end}.
    The blue (green) dotted lines represent Higgs (Leggett) propagators, the wavy black line represents the electromagnetic field and the solid black line the Nambu Greens function.
    (a,b) Higgs and phase susceptibilities $\chi_i^{\sigma_1\sigma_1}$, $\chi_i^{\sigma_3\sigma_3}$. 
    (c) Coupling of Higgs modes where vertex is the inverse of Eq.~\eqref{eqn:v}. (d) Josephson coupling
    of phase modes responsible for Leggett mode. The coupling matrix $J$ is defined in Eq.~\eqref{eq:leggettmatrix}. (e) Paramagnetic coupling of Higgs modes with susceptibility 
    $\chi^{\sigma_0\sigma_0\sigma_1}$. 
    (f) Paramagnetic coupling of phase modes with $\chi^{\sigma_3\sigma_3}$. 
    Other couplings at Gaussian level vanish in the presence of particle-hole symmetry.  }
    \label{fig:diagrams-higgs-leggett}
\end{figure}
\begin{figure}[ht]
    \centering
    \includegraphics[width=0.9\columnwidth]{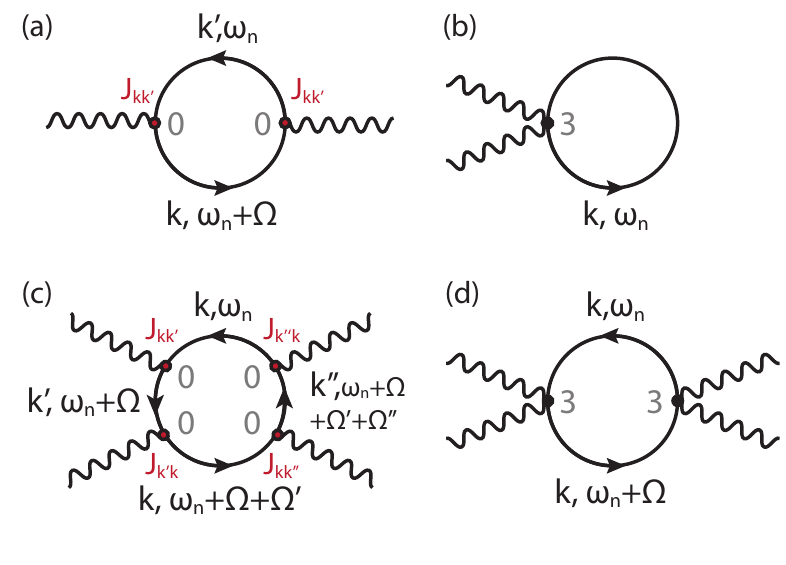}
    \caption{Diagrammatic representation of quasiparticle terms in the effective action $S[\Delta_i,\theta_i,\mathbf{A}]$ Eq.~\eqref{eq:seffappdx}. 
    Paramagnetic (a) and diamagnetic (b) terms defining the linear response current $\mathbf{j}\big|_1$. 
    The paramagnetic contribution (a)
    vanishes in the clean limit.
    Paramagnetic (c) and diamagnetic (d) terms contributing to nonlinear current $\mathbf{j}\big|_3$. }
    \label{fig:diagrams-j}
\end{figure}
\begin{figure}[ht]
    \centering
    \includegraphics[width=\columnwidth]{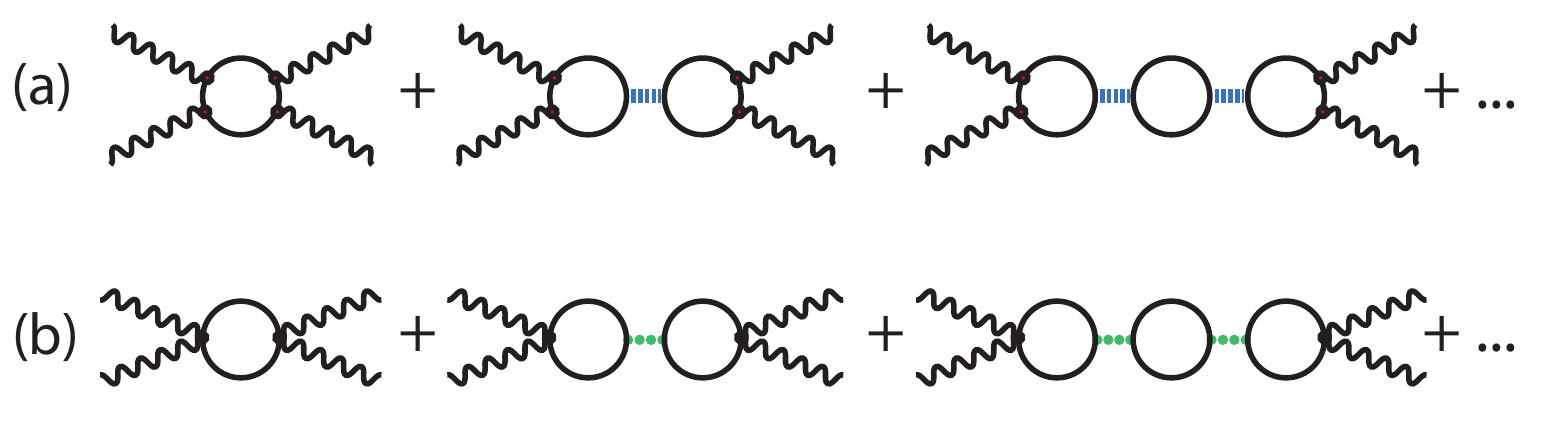}
    \caption{RPA summation of collective fields in the effective action. 
    (a) Higgs modes renormalize paramagnetic non-linear current. Here, blue dashed lines correspond to the coupling matrix $U/2$.
    (b) Phase modes renormalize diamagnetic current. Green dotted lines denote coupling matrix $J^{-1}$. }
    \label{fig:diagrams-RPA}
\end{figure}
We first present a perturbative solution of above Hamiltonian by a path-integral formalism in 
imaginary time $\tau$ \cite{VanOtterlo1999, Sharapov2002, Benfatto2004, Cea2018}. The full problem is formally captured by the partition function
$\mathcal{Z} = \int\mathcal{D}(c^\dagger c) e^{-S}$
with the euclidean action
\begin{equation}
    S = \int_{0}^{\beta}d\tau
	\left( 
		\sum_{i\mathbf{k}\sigma}^{} 
		c_{i\mathbf{k}\sigma}^\dagger  \partial_\tau
		c_{i\mathbf{k}\sigma} 
		+ \mathcal{H}
	\right)\,.
\end{equation}
As detailed in Appendix~\ref{apdx:effective-action}, we decouple the interacting part of $\mathcal{H}$ in the paring channel, introducing collective fields $\Delta_i(\omega_n)\exp\left(i\theta(\omega_n) \right)$. $\Delta_i$ and $\theta_i$ describe amplitude and phase fluctuations, respectively,
of the superconducting condensate. In the present work we restrict ourselves to collective fluctuations 
in time only, i.e. we focus on $\mathbf{k}=0$ excitations of Higgs and phase fields.

Performing the fermionic path integral results in an effective action $S[\Delta_i,\theta_i,\mathbf{A}]$ in bosonic and classical EM fields (see Eq.~\eqref{eq:seffappdx}),
where now $\mathcal{Z} = \int\prod_i\mathcal{D} \Delta_i \mathcal{D}_i \theta_i e^{-S[\Delta_i,\theta_i,\mathbf{A}]}$. 
We only keep terms quadratic in collective fields $\Delta_i, \theta_i$ and to fourth order in $\mathbf{A}$.
The resulting terms are diagrammatically presented in Fig.~\ref{fig:diagrams-higgs-leggett} and Fig.~\ref{fig:diagrams-j}
and their integral expressions are derived in Appendix~\ref{apdx:effective-action}.

The diagrammatic representation contains Higgs fields $\Delta_i(\omega)$ (blue-dashed lines), 
phase fields (green-dotted lines), and EM fields (wavy lines). Paramagnetic coupling to the photon field
corresponds to vertices with a single photon field line, implying the factor $A(\omega)$.
Diamagnetic vertices with two photon field lines contribute the term $A^2(\omega) = \int d\omega' A(\omega-\omega')A(\omega')$.
Only paramagnetic vertices introduce external momentum.
Solid black lines correspond to mean-field Nambu Green's functions
\begin{equation}
    G_{0,i}=\left[ i\omega_n -\epsilon_{i\mathbf{k}} \sigma_3 + \Delta_i \sigma_1 \right]^{-1}
\end{equation}
and loops imply a trace over Nambu indices,
frequencies, and momenta.

Figs.~\ref{fig:diagrams-higgs-leggett} and \ref{fig:diagrams-j} are a complete representation of 
all terms in the quadratic action in the presence of particle-hole symmetry and 
impurities in the MB approximation. 

In the clean limit paramagnetic photon lines no longer carry momentum and, as a consequence,
diagrams \ref{fig:diagrams-higgs-leggett}(e) and \ref{fig:diagrams-j}(a) vanish. The inclusion of paramagnetic
diagrams with vertices $J_{i\mathbf{kk'}}$ determined by the MB model 
is the main difference of the diagramatic formalism from other literature \cite{2016Benfatto-Leggett,Cea2016}.

Absence of diagram \ref{fig:diagrams-higgs-leggett}(e) in the clean limit implies that the Higgs 
mode does not couple to light without impurities.
However, when a non-parabolicity of the bandstructure is taken into account,
a diamagnetic coupling to the Higgs mode arises, yielding an additional, non-vanishing diagram \cite{Tsuji2015,Cea2016,Murotani2019}.

We note that paramagnetic and diamagnetic terms do not mix in the present model. 
Consequently, the partition function factors into 
two contributions $\mathcal{Z}=\mathcal{Z}_{\text{para}}\mathcal{Z}_{\text{dia}}$. 
Since only the paramagnetic part is affected by impurities, and since $\mathcal{Z}_{\text{para}}$ does not contain 
phase contributions, we conclude that only the Higgs mode and quasiparticles are sensitive to impurity scattering
in the MB approximation.

The path integrals over $\Delta_i, \theta_i$ can be performed exactly at Gaussian level. This is equivalent to 
an RPA renormalization of the quasiparticle terms 
diagrammatically represented in Fig.~\ref{fig:diagrams-RPA} where the dashed and dotted lines correspond to
coupling matrices $U_{ij}/2$ and Josephson coupling matrices $J^{-1}_{ij}$, respectively.
One is left with $S[A(\omega)]$, explicitly given in Eq.~(\ref{eq:effaction}).
A functional derivative with respect to $A(\omega)$ gives the current
\begin{equation}
    j(\omega) = - \frac{\delta S[A(\omega)]}{\delta A(\omega)} \,.
    \label{eqn:fderiv}
\end{equation}

\subsection{Density matrix equations of motion}
We solve for the time dynamics of above Hamiltonian using a density matrix
approach. To this end, we define the density matrix $\rho =
\ket{\psi_0}\bra{\psi_0}$, or, in the basis of Bogoliubov–de Gennes,
\begin{align}
    \VV{\psi_{i\mathbf{k}}^1}{\psi_{i\mathbf{k}}^2} = \MM{u_{i \mathbf{k} }}{-v_{i \mathbf{k}}}{v^*_{i \mathbf{k}}}{u_{i \mathbf{k}}} \VV{c_{i \mathbf{k} \uparrow}}{c^\dagger_{i (-\mathbf{k}) \downarrow}}\,,
\end{align}
we have
\begin{eqnarray}
  \rho =
  \begin{pmatrix}
    \rho_{i\mathbf{kk'}}^{11} &
    \rho_{i\mathbf{kk'}}^{12} \\
    \rho_{i\mathbf{kk'}}^{21} &
    \rho_{i\mathbf{kk'}}^{22}
  \end{pmatrix}
  =
  \begin{pmatrix}
    \langle \psi^{1\dagger}_{i\mathbf{k}} \psi^{1}_{i\mathbf{k'}}\rangle
    &
    \langle \psi^{1\dagger}_{i\mathbf{k}} \psi^{2}_{i\mathbf{k'}} \rangle
    \\
    \langle 
    \psi^{2\dagger}_{i\mathbf{k}} \psi^{1}_{i\mathbf{k'}}
    \rangle
    &
    \langle 
    \psi^{2\dagger}_{i\mathbf{k}} \psi^{2}_{i\mathbf{k'}} 
    \rangle
  \end{pmatrix} 
  \,.
\end{eqnarray}
The time dependence of $\rho$ is given by Heisenberg's equation of motion,
\begin{eqnarray}
  i\partial_t \rho = \left[ \rho, H \right]\,,
\end{eqnarray}
where $H$ is the operator $\mathcal{H}$ in the BdG basis. 

We are interested in computing the dynamics of the current
$ \mathbf{j} = -\big\langle \frac{\delta \mathcal{H}}{\delta
\mathbf{A}}\big\rangle =
\mathbf{j}_P + \mathbf{j}_D$, consisting of a paramagnetic and diamagnetic
contribution,
\begin{eqnarray}
  \mathbf{j}_P 
	&=&
  	\sum_{i\mathbf{kk'}}^{}
	\mathbf{J}_{i\mathbf{kk'}\sigma}\,
	\langle
	c_{i\mathbf{k}\sigma}^\dagger c_{i\mathbf{k'}\sigma}
	\rangle\,,
	\\
  \mathbf{j}_D &=& -\sum_{i\mathbf{k}\sigma}^{} \frac{s_i e^2}{m_i} 
  \mathbf{A} \,
	\langle
	c_{i\mathbf{k}\sigma}^\dagger c_{i\mathbf{k}\sigma}
	\rangle\,,
\end{eqnarray}
as well as the dynamics of the superconducting order parameter
\begin{eqnarray}
  \Delta_i = \sum_{j\mathbf{k}}^{}U_{ij} 
  \langle c_{j(-\mathbf{k})\downarrow}c_{j\mathbf{k}\uparrow}\rangle \,.
\end{eqnarray}
To apply the MB substitution, we further expand above equations of motion in
orders of $A(t)$. To account for effects of a THG response, we consider terms up to
third order. As detailed in Appendix~\ref{apdx:model}, the current only has odd order
components $\mathbf{j} = \mathbf{j}\big|_0 +\mathbf{j}\big|_3 + \dots$ and the
gap contains even contributions of $A$, $\Delta = \Delta\big|_0 +
\delta\Delta\big|_2 +\dots$. 

Finally, we exploit the rotational invariance of our model and perform the
integral over angular degrees of freedom explicitly. Thus, by replacing all momentum summations
by an integral 
$\sum_{\mathbf{k}}^{} \rightarrow
N_i(0)\int_{}^{}d\epsilon_{i\mathbf{k}}\int_{}^{}
\frac{d\Omega_{\mathbf{k}}}{4\pi}$, we effectively reduce the model to a
one-dimensional system. Note that rotational invariance of our continuum model
neglects polarization dependence of observable quantities.

We are left to compute the equations of motion of the first order quasiparticle expectation values, $\rho_{i\mathbf{kk'}}\big|_1$, and the
angle-averaged quantities
\begin{eqnarray}
  R^{ab}_i(\epsilon_{i|\mathbf{k}|},\epsilon_{i|\mathbf{k}'|}) &=& 
	\frac{1}{\int \frac{d \Omega_\mathbf{k}}{4 \pi} \frac{d \Omega_\mathbf{k'}}{4 \pi} \left| \mathbf{J}_{i\mathbf{kk'}} \cdot \mathbf{e} \right|^2 }
	\\
		\nonumber
	&& \times
	\int \frac{d \Omega_\mathbf{k}}{4 \pi} \frac{d \Omega_\mathbf{k'}}{4 \pi}
	\mathbf{e} \cdot \mathbf{J}_{i\mathbf{kk'}}	
	\rho_{i\mathbf{kk'}}^{ab}\big\rvert_3\,,
  \\
 r_i^{ab}(\epsilon_{i|\mathbf{k}|}) &=& \int \frac{d \Omega_\textbf{k}}{4 \pi} \rho_{i\mathbf{kk}}^{ab}\big|_2 \,.
\end{eqnarray}
We solve them numerically using a Runge-Kutta solver on a discretized energy grid
$\epsilon_{|\mathbf{k}_i|}$ of up to $10^3$ points in the interval $\left[
  -\omega_D, \omega_D
\right]$.
A detailed derivation and explicit presentation of the full equations of
motion is given in Appendix~\ref{apdx:model}.

\section{Single-band superconductivity}
\label{sec:singleband}

Motivated by the experiment of Matsunaga \textit{et al.} \cite{Matsunaga2013} we choose parameters
$\Delta=\SI{1.3}{\milli\electronvolt}$,
$\epsilon_{F}=\SI{1}{\electronvolt}$, $m=0.78 m_e$, $s=1$, $\omega_D =
\SI{20}{\milli\electronvolt}$ that reflect
measurements and ab-initio calculations on NbN \cite{Babu2019}.

\subsection{Optical conductivity}
\label{sec:oc}
We begin by computing the optical conductivity in linear response,
\begin{eqnarray}
  \sigma(\omega)=\frac{j(\omega)\big|_1}{i\omega A(\omega)} \,.
  \label{eqn:sigma1}
\end{eqnarray}
This can be done in either of two ways. First, by implementing a time-dependent density matrix simulation
with pulse $A(t)$. The numerically evaluated current $j(t)\big|_1$ and the pulse are then Fast-Fourier transformed
and Eq.~(\ref{eqn:sigma1}) is evaluated. Here, one needs to choose a pulse of sufficient $\omega$-bandwidth
such that the region of interest is covered. 

The second way involves the functional derivative of the
diagrams in Fig.~\ref{fig:diagrams-j}(a,b) according to Eq.~(\ref{eqn:fderiv}). At $T=0$
one obtains the expression for the real part
\begin{eqnarray}
\sigma'(\omega) &=&\frac{1}{i\omega}\frac{v_{F}^2}{3N}  \int d\epsilon d\epsilon' W(\epsilon,\epsilon') \chi''^{\sigma_0\sigma_0}(\epsilon,\epsilon',\omega)
\nonumber
\\
&=&
\frac{1}{i\omega}\frac{v_{F}^2N}{3} 
\int d\epsilon d\epsilon' W(\epsilon,\epsilon')\left(1-\frac{\epsilon\epsilon'+\Delta^2}{EE'}\right)
\nonumber
\\
&&\quad\quad\quad\quad\quad\quad \times
\frac{E+E'}{(w+i\eta)^2-(E+E')^2}
\label{eqn:sigma-expl}
\end{eqnarray}
where $E'=\sqrt{\Delta^2+\epsilon'^2}$, $W(\epsilon,\epsilon')$ is the Lorentzian 
of Eq.~(\ref{eq:W-lorentz}), N the density of states at the Fermi surface, and $\eta$ is an infinitesimal positive constant.

\begin{figure}
    \centering
    \includegraphics[width=\columnwidth]{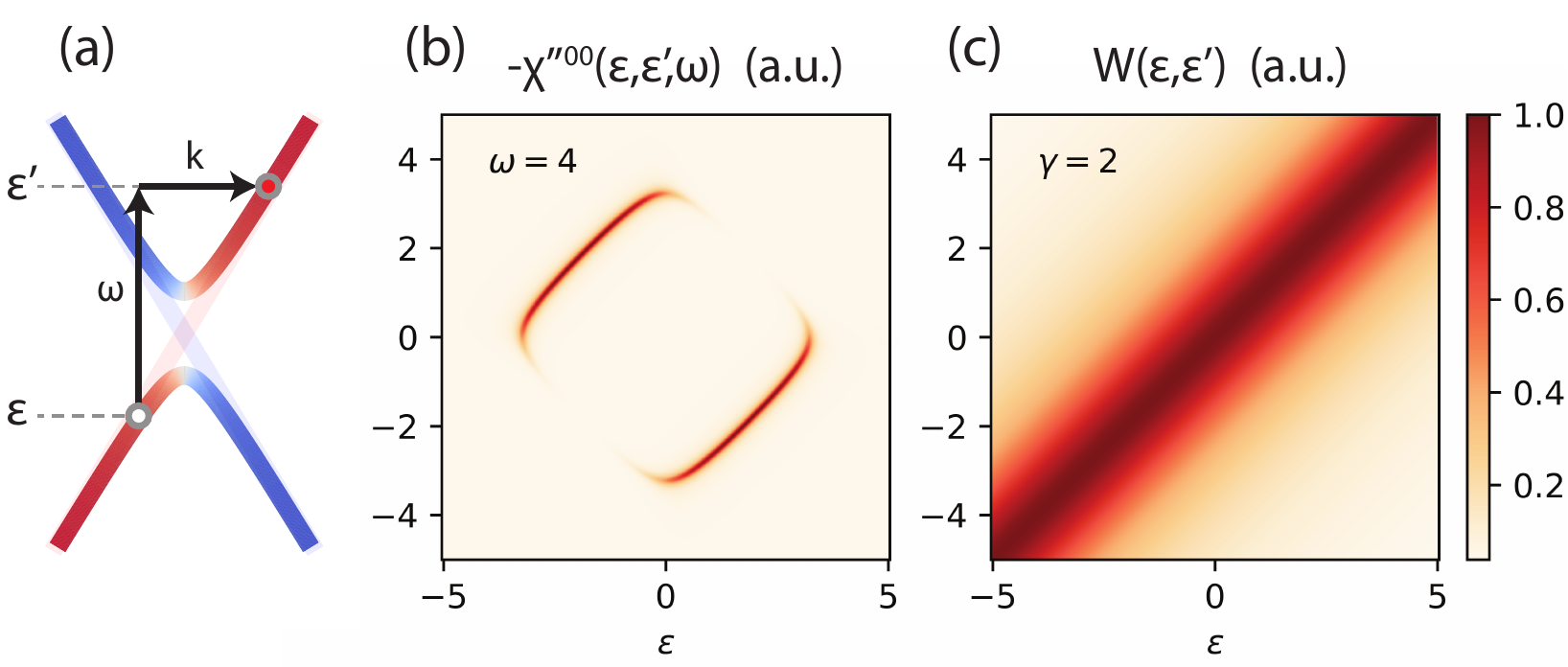}
    \caption{(a) Schematic of a particle-hole excitation process where the pulse contributes energy $\omega$
    and momentum $k$. Red (blue) colors indicate electron (hole) character. 
    (b) The susceptibility $-\chi''^{\sigma_0\sigma_0}$ has finite weight only 
    for $\epsilon,\epsilon'$ corresponding to valid state in an excitation process
    with $\omega=4$.
    Rounded features are a result of the gap $2\Delta$.
    For $\omega<2\Delta$, $\chi''^{\sigma_0\sigma_0}$ is identically zero since no optical excitation is not possible.
    (c) Momentum conservation is enforced by the factor $W(\epsilon-\epsilon')$ in Eq.~\eqref{eqn:sigma-expl}.}
    \label{fig:x00}
\end{figure}

We can understand the analytical structure of $\sigma'(\omega)$ by inspecting the susceptibility $\chi''^{\sigma_0\sigma_0}(\epsilon,\epsilon',\omega)$.
For $\omega<2\Delta$ it vanishes exactly. For $\omega>2\Delta$ its structure is exemplary shown in Fig.~\ref{fig:x00}.
We observe two straight spectral lines at $\epsilon'=\pm \omega+\epsilon$.
These features can be understood in the picture of a particle-hole or hole-particle excitation process,
illustrated in Fig.~\ref{fig:x00}(a). $\chi''^{\sigma_0\sigma_0}(\epsilon,\epsilon',\omega)$ has non-zero
spectral weight at given $\epsilon,\epsilon'$ if an occupied state at $\epsilon$ can be excited into 
a state at $\epsilon'$ by a photon of frequency $\omega$. Multiplication of the integrand in Eq.~(\ref{eqn:sigma-expl})
with $W(\epsilon-\epsilon')$ enforces momentum conservation.

In this picture it is easy
to see that the total spectral weight $\chi'^{\sigma_0\sigma_0}(\omega)=\int d\epsilon d\epsilon' \chi'^{\sigma_0\sigma_0}(\epsilon,\epsilon',\omega)$ 
should be approximately proportional to $\Theta(\omega-2\Delta) (\omega-2\Delta)$, where $\Theta$ is the Heaviside function.  Since $W(\epsilon-\epsilon')$ is constant along contours $\epsilon'=\pm \omega +\epsilon$,
we find the simple analytical approximation
\begin{equation}
    \sigma'(\omega) \propto \Theta(\omega-2\Delta) (\omega-2\Delta) \frac{\gamma}{\omega^2 + \gamma^2} 
    \label{eqn:sigma-analutical}
\end{equation}
that holds for $\omega \gg 2 \Delta$ in the dirty limit $\gamma \gg \Delta$.

\begin{figure}[t]
    \centering
    \includegraphics[width=\columnwidth]{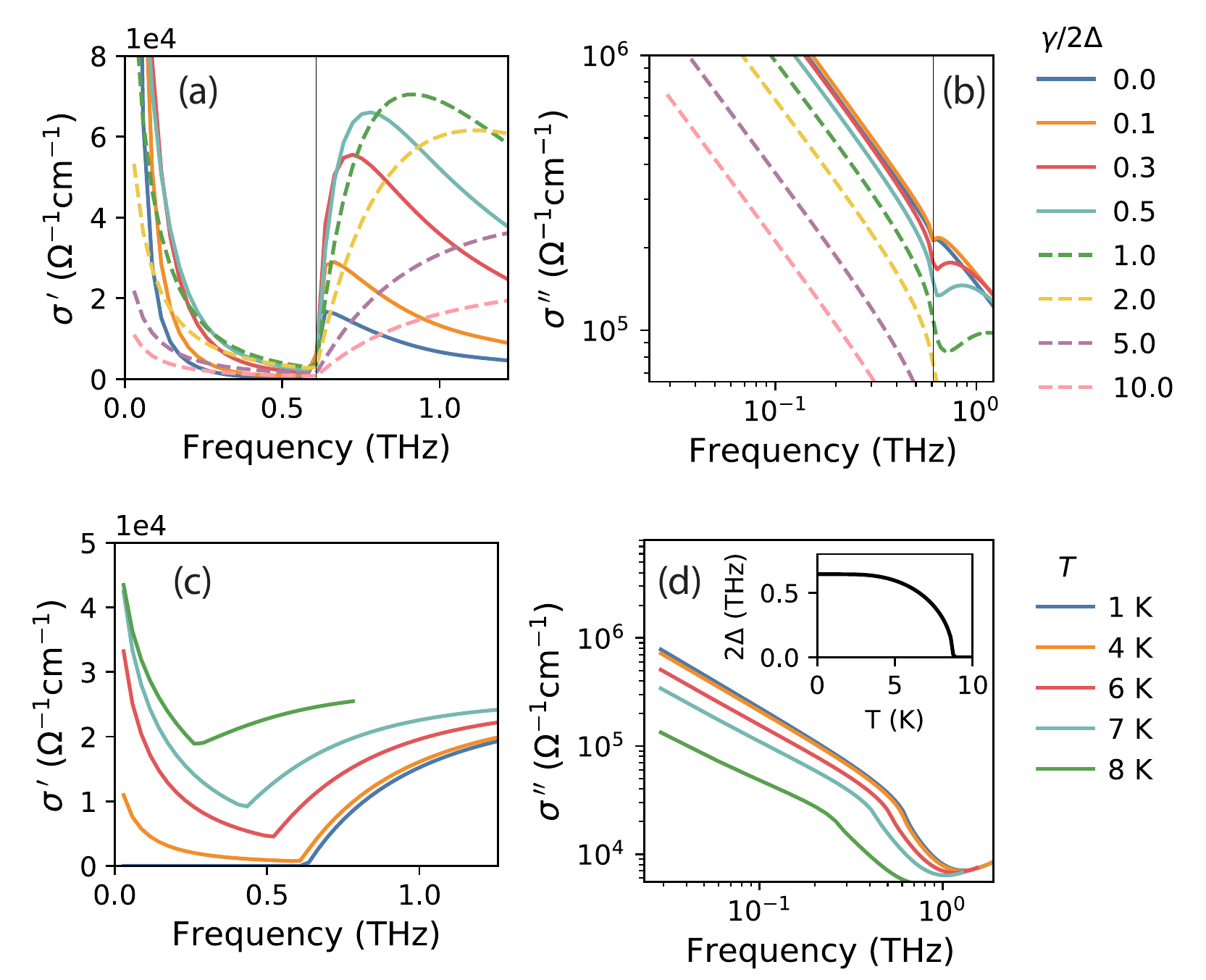}
    \caption{Real part $\sigma'$ and imaginary part $\sigma''$ of the optical
    conductivity to first order in the vector potential $A$.
    (a),(b) Impurity scattering rates dependence for fixed temperature $T=\SI{4}{\kelvin}$.
    (c),(d) Temperature dependence for fixed scattering rate $\gamma/2\Delta=10$.
    $\sigma'$ shows a characteristic conductivity gap below $T_C$
    and both $\sigma'$, $\sigma''$ diverge in the static limit.
    The inset in (d) shows the temperature dependence of the gap.}
\label{fig:cond-lin}
\end{figure}

In Fig.~\ref{fig:cond-lin} we plot numerically evaluated real and
imaginary parts $\sigma'(\omega),\sigma''(\omega)$ of the optical conductivity 
for various impurity concentrations and temperatures. 
$\sigma'$ shows a clear conductivity gap below $2\Delta$. 
In the clean limit, a pronounced coherence peak is observed around $2\Delta$,
reflecting the additional density of states amassed above the quasiparticle gap. 
The conductivity peak grows and shifts to higher $\omega$ as $\gamma$ is increased. 
It then broadens
into the characteristic dome shape frequently observed in experiment \cite{Matsunaga2013,Mattis1958, Zimmermann1991}. 
In the $T\rightarrow 0$ limit, the
conductivity is expected to show a condensate $\delta$-peak at $\omega=0$
which is not numerically resolveable. 
Instead, we observe a buildup of spectral weight around $\omega=0$
as the condensate peak is broadened at finite temperatures.
The imaginary part $\sigma''$ follows a $1/\omega$ power law as expected for a 
superconducting state.

The linear response optical conductivity contains information of the bandstructure
only and is unaffected by collective modes. This can be inferred from the diagrammatic description where
all terms in the RPA renormalization of diagram Fig.~\ref{fig:diagrams-j}(a) 
containing $\mathbf{k}=0$ collective fluctuations vanish exactly. 
To reveal the presence of collective modes, we turn to the dynamics of the superconducting
order parameter and the non-linear current $j\big|_3$ and additionally model
realistic THz pulses in a pump-probe setting.

\subsection{Excitation of Higgs mode}

\begin{figure}[t]
	\centering
	\includegraphics[width=\columnwidth]{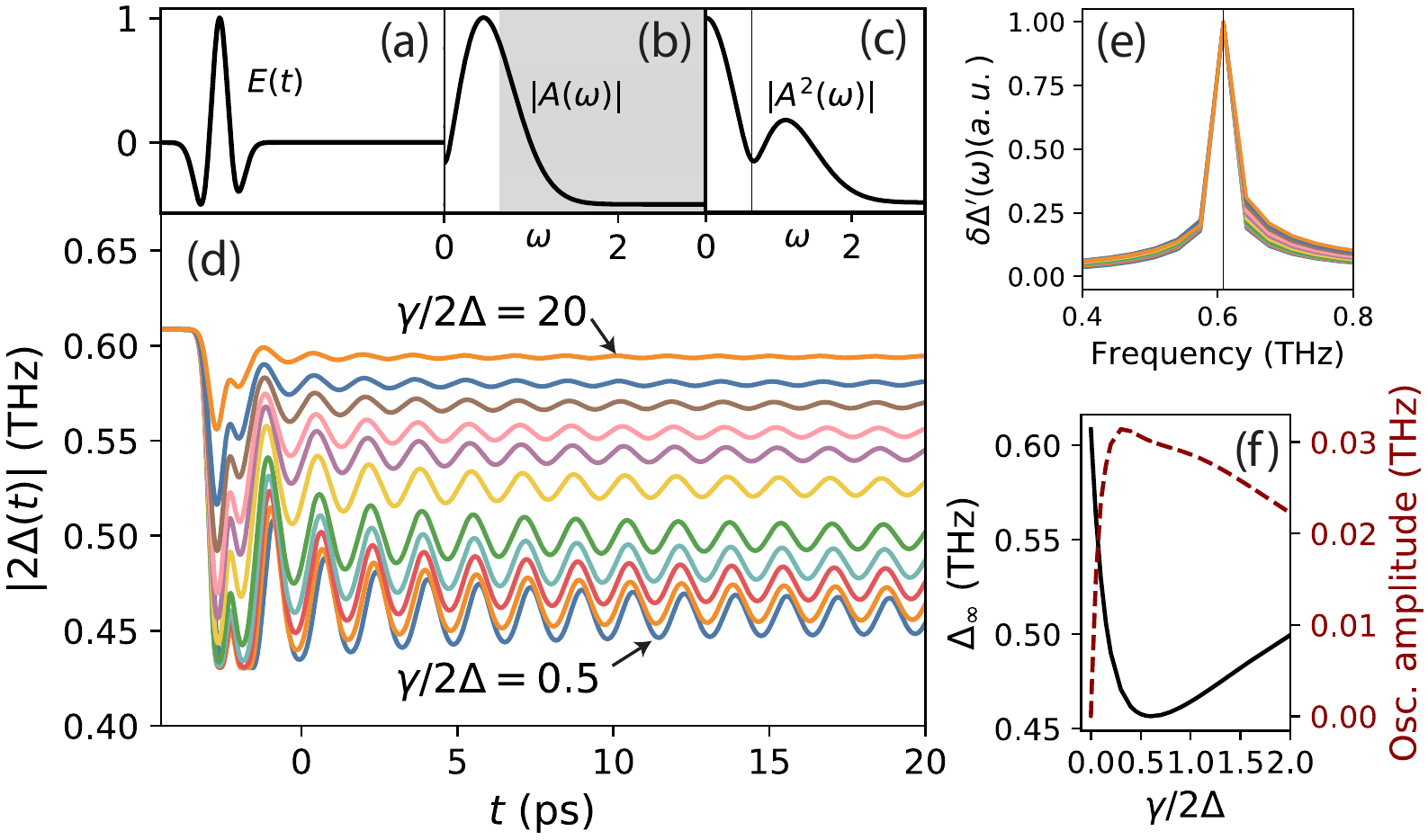}
	\caption{(a) Pulse field $E(t)$ realizing a quench. (b) Spectral
	composition $|A(\omega)|$.
	The gray shaded area illustrates the
quasi-particle continuum. (c) Spectral composition $|A^2(\omega)|=|\int d\omega' A(\omega-\omega')A(\omega')|$ of the second
order component $A^2(t)$ responsible for excitation of collective modes. The
peak around zero frequency corresponds to a DFG process while the peak at finite
$\SI{1.2}{\tera\hertz}$ is a SFG process. (d) Evolution of the magnitude of the
order parameter $\left|2\Delta(t)\right|$ for impurity strength varying from
$\gamma/2\Delta=0.5$ to $20$ and Fourier spectrum of the gap oscillations (e).
(f) Relaxation value $\Delta_{\infty}$ and amplitude of oscillation show a very
similar dependence as a function of disorder strength which has maximum effect
at around $\gamma \approx \Delta$.}
\label{fig:singleband-gap}
\end{figure}

We choose the electromagnetic pulse form $A(t) = A_0 \exp\left( -(t-t')^2/2\tau^2
\right)\cos \Omega t$ with coefficients to match the reported data of Ref.~\cite{Matsunaga2013}. The resulting
 waveform is shown in Fig.~\ref{fig:singleband-gap}(a).

A characteristic property of a pump pulse is its pulse 
length $\tau$ compared to the natural
timescale of the superconductor $1/\Delta$. For $\tau \ll 1 / \Delta$
the superconductor is \textit{quenched}, while it is \textit{adiabatically
driven} in the
opposite limit of $\tau \gg  1 / \Delta$. 

\begin{figure}[t]
    \centering
    \includegraphics[width=\columnwidth]{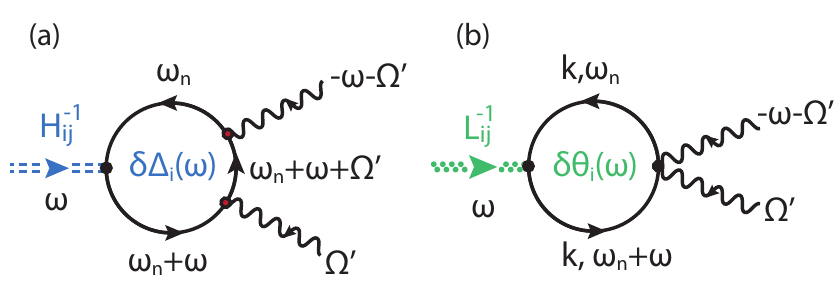}
    \caption{Diagrammatic representation of (a) $\delta\Delta_i(\omega)$ and (b) $\delta\theta_i(\omega)$. Double lines correspond to the RPA summation of Fig.~\ref{fig:diagrams-RPA}.}
    \label{fig:diagrams-dd}
\end{figure}

The different behavior in the two limits can be intuitively understood 
within the diagrammatic picture. Here, the pulse induced
change of the order parameter $\delta\Delta(\omega)$ is given by the diagram in Fig.~\ref{fig:diagrams-dd}(a)
which has the integral expression
\begin{align}
    \delta \Delta(\omega) &= \frac{1}{2} \int d\omega'
    \sum_{\mathbf{kk'}}
    \left| J_{\mathbf{kk'}}\right|
    \frac{\chi^{\sigma_0\sigma_0 \sigma_1}(\omega,\omega',\mathbf{k,k'})}{\chi^{\sigma_1\sigma_1}(\omega)+2/U}
    \nonumber
    \\ 
    &\,\times
    A(\omega')A(-\omega-\omega') \,.
    \label{eqn:dd-sngle}
\end{align}
Presence of a collective Higgs mode
translates into a peak of the kernel $K(\omega) = \left(\chi^{\sigma_1\sigma_1}(\omega)+2/U\right)^{-1}$
at the characteristic mode energy
$\omega_H=2\Delta$. Excitation of the collective mode, 
however, is only possible if energy conservation is satisfied, i.e. if 
$A(\omega')A(-\omega_H-\omega')$ is finite for some $\omega'$.
Higgs oscillations are therefore expected when the Fourier transform of the squared vector
potential
$A^2(\omega)=\int d\omega' A(\omega-\omega')A(\omega')$ overlaps with the mode-energy $\omega_H$. 
The double-peaked structure of $A(\omega)$ is shown in Fig.~\ref{fig:singleband-gap}(c). The first peak,
centered at $\omega=0$, corresponds to a difference frequency generation process (DFG),
while the second peak at $\omega=2\Omega$ corresponds to a sum frequency generation process
(SFG). The resonance frequency of the Higgs mode, $\omega_H$,
is illustrated by a vertical line. Remaining terms in Eq.~\eqref{eqn:dd-sngle}
describe the coupling to light in presence of impurities 
and ensure momentum conservation in a virtual two step excitation process.

Let us now consider two limiting cases of the optical pulse width. 
For $\Delta\tau\ll 1$, the frequency spectrum of $A^2(\omega)$ is very
broad. The response of $\delta\Delta(\omega)$ is then dominated by the sharp
resonance peak of $K(\omega)$ giving rise to pronounced $2\Delta$-oscillations
of the superconducting gap in the time domain. 
Since the DFG peak is 
guaranteed to overlap with the Higgs resonance,
these oscillations will always be present,
independent of the frequency of the optical pulse. The SFG process only contributes
if the pulse frequency lies in the vicinity of $\Omega \approx \Delta$.

In the transient limit, $\Delta \tau\gg 1$, the spectrum of
$\delta\Delta(\omega)$ is finite only for a narrow region around $2\Omega$. 
In the time-domain, the gap
shows forced $2\Omega$-oscillations which are
resonantly enhanced for $2\Omega\approx2\Delta$.

Following Matsunaga \cite{Matsunaga2013}, we choose a pulse with $\Delta \tau = 0.68$, closest to the
quench scenario, and perform simulations within the density-matrix formulation.
The order parameter responds to the THz pulse
by a marked drop followed by damped oscillations around a new asymptotic value
$\Delta_{\infty}=\Delta(t\rightarrow \infty)$ of frequency $2\Delta=\SI{0.6}{\tera\hertz}$ as displayed in Figs.~\ref{fig:singleband-gap}(d-e).
The drop of the equilibrium gap is captured by the $\omega=0$ component of $\delta \Delta$.
Evaluating Eq.~\eqref{eqn:dd-sngle} for $\omega=0$, one finds that $\chi^{\sigma_0\sigma_0 \sigma_1}(\omega=0,\omega',\mathbf{k,k'})$
is finite only for $\omega'>2\Delta$, similar to the discussion in Sec.~\ref{sec:oc}. 
Consequently, $\delta\Delta(0)$ is non-zero only if $|A(\omega)|^2$ overlaps with the quasiparticle continuum, which is
illustrated in Fig.~\ref{fig:singleband-gap}(b). In physical terms, depletion of the 
superconducting order parameter is a consequence of quasiparticle excitation by $A(\omega)$.

Both the oscillation amplitude and $\Delta_{\infty}$ show a
strong dependence on the impurity scattering rate and are peaked at $\gamma\approx
\Delta$ as shown in Fig.~\ref{fig:singleband-gap}(f). This is a consequence of
momentum conservation. For $\gamma\rightarrow 0$,
 Higgs oscillations vanish exactly.

We note that order parameter dynamics are expected to show oscillations of frequency 
$2\Delta_\infty$ and not, as in our case, $2\Delta(t=0)$ \cite{PhysRevLett.96.230404,Krull2014}. $2\Delta_\infty$ oscillations
have also been observed in experiment \cite{Matsunaga2013}. The discrepancy can be attributed
to the expansion in powers of the pump field $A(t)$ performed in the time-dependent density matrix formalism. 
If contributions to $\delta\Delta$ beyond the second order
are considered, the oscillation frequency of the order parameter should correctly reflect the non-equilibrium value
$2\Delta_{\infty}$. Strictly speaking, our model is fully valid only in the limit of small pump fields
where the difference between $\Delta$ and $\Delta_{\infty}$ is negligible.

\subsection{Pump-probe spectroscopy}
\begin{figure}[t]
	\centering
	\includegraphics[width=\columnwidth]{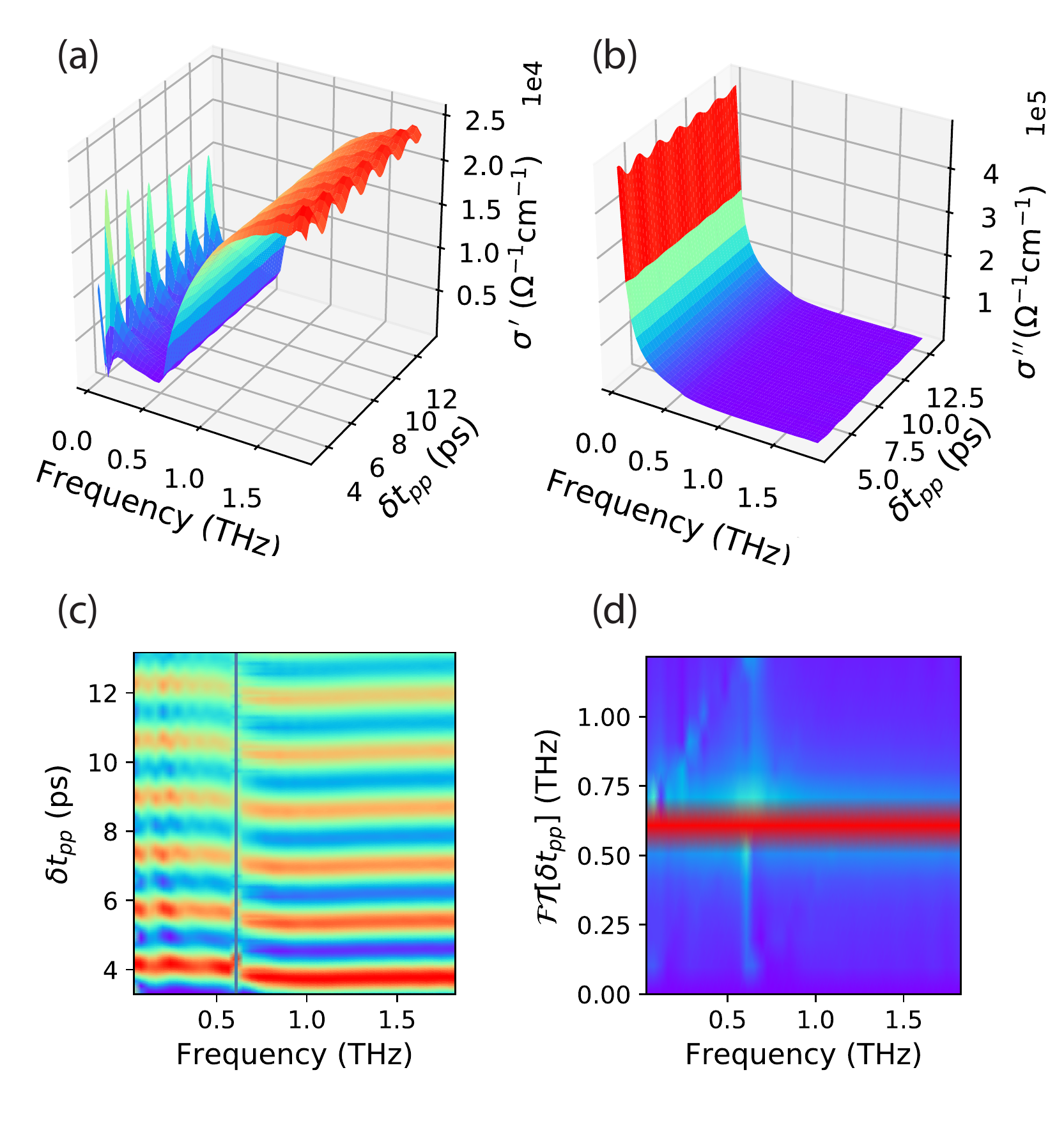}
	\caption{(a,b) Real and imaginary part of conductivity spectra for sweeped pump-probe delay $\delta
	t_{\text{pp}}$ to third order in $A$. (c) False-color plot of
$\sigma' $ which was average-subtracted and normalized to show the oscillations.
A phase shift occurs across the resonance at $2\Delta$ of the quench pulse
frequency. (d) Fourier spectrum of panel (c) showing that frequency of
conductivity oscillation is peaked at $2\Delta$.}
\label{fig:cond-pp}
\end{figure}

Higher orders of the optical conductivity include
contributions of collective modes that smooth out the absorption
edge and add spectral weight inside the conductivity gap.
Here, we calculate the non-linear contribution,
\begin{align}
    \sigma(\omega, \delta t_{pp})=\frac{j(\omega)\big|_1+j(\omega)\big|_3}{i\omega
A(\omega)} \,,
\end{align}
in a pump-probe setting of the time-dependent density-matrix formalism.
To this end, we pump the system with an intense 
pulse of fluence $A_0=0.5 \times 10^{-8}\si{\joule\second\per\coulomb\per\meter}$ and, after a delay $\delta t_{pp}$, apply a weak probe
pulse. Following experimental schemes \cite{Matsunaga2012}, we perform two simulations.
First, we simulate
both a pump and a probe pulse to compute $j_{pp}$. In a second simulation we apply the pump only, obtaining $j_{p}$. 
We then compute the optical conductivity
from the difference in currents $j=j_{pp}-j_{p}$. This ensures that resilient
contributions of the pump do not affect the optical conductivity.

Figs.~\ref{fig:cond-pp}(a-b) show the real and imaginary part of the optical
conductivity $\sigma(\omega,\delta t_{pp})$ as a function of frequency and
pump-probe delay. It can be seen that the third-order contribution
$j\big|_3$ adds spectral weight to the conductivity below
absorption gap. The conductivity shows clear oscillations in $\delta t_{pp}$, 
as emphasized in Fig.~\ref{fig:cond-pp}(c) where
the dome-shaped envelope has been subtracted and the remaining signal was normalized
for each $\omega$. A Fourier transform of these
oscillations, shown in Fig.~\ref{fig:cond-pp}(e), indicates that the
oscillation frequency matches the resonance frequency of the
Higgs mode $2\Delta$. 

Our results show that signatures of the Higgs mode are measurable in the
pump-probe response of the optical conductivity. Yet, to excite the Higgs mode,
impurities are crucial. We find that the calculated time-resolved optical response of a single-band
superconductor in the dirty-limit is in good agreement with the experimentally measured
response \cite{Matsunaga2013}.

\section{Multi-band superconductivity}
\label{sec:multiband}

Motivated by the good agreement of the theory with experimental data for a single-band superconductor, 
we now
turn to the case of a two-band superconductor. For concreteness, we focus 
on the superconducting state
of MgB$_2$. We model the
$\pi$- and $\sigma$-bands believed to be responsible for superconductivity by
choosing material parameters 
$\Delta_{\pi}=\SI{3}{\milli\electronvolt}$,
$\Delta_{\sigma}=\SI{7}{\milli\electronvolt}$, 
$\epsilon_{F,\pi}=\SI{2.9}{\electronvolt}$,
$\epsilon_{F,\sigma}=\SI{0.7}{\electronvolt}$, 
$m_\pi=0.85m_e$,
$m_\sigma=1.38 m_e$, 
$\omega_D=\SI{50}{\milli\electronvolt}$,
$s_\pi=1$,
$s_\sigma=-1$ \cite{Kortus2001}.

Convincing evidence for the two-band character of MgB$_2$ has been found in tunneling measurements \cite{Giubileo2001,Ivarone2002} 
and ARPES \cite{Tsuda2003}.
However, optical linear response probes have only revealed signatures of a superconducting gap in the
$\pi$-band \cite{Kaindl2001, Kovalev2020}. A recent work \cite{Kovalev2020} on third harmonic generation suggests strong evidence
of a collective Higgs resonance in the $\pi$-band, but no collective response in the $\sigma$-band was observed.

\subsection{Optical conductivity}

\begin{figure}[t]
	\centering
	\includegraphics[width=\columnwidth]{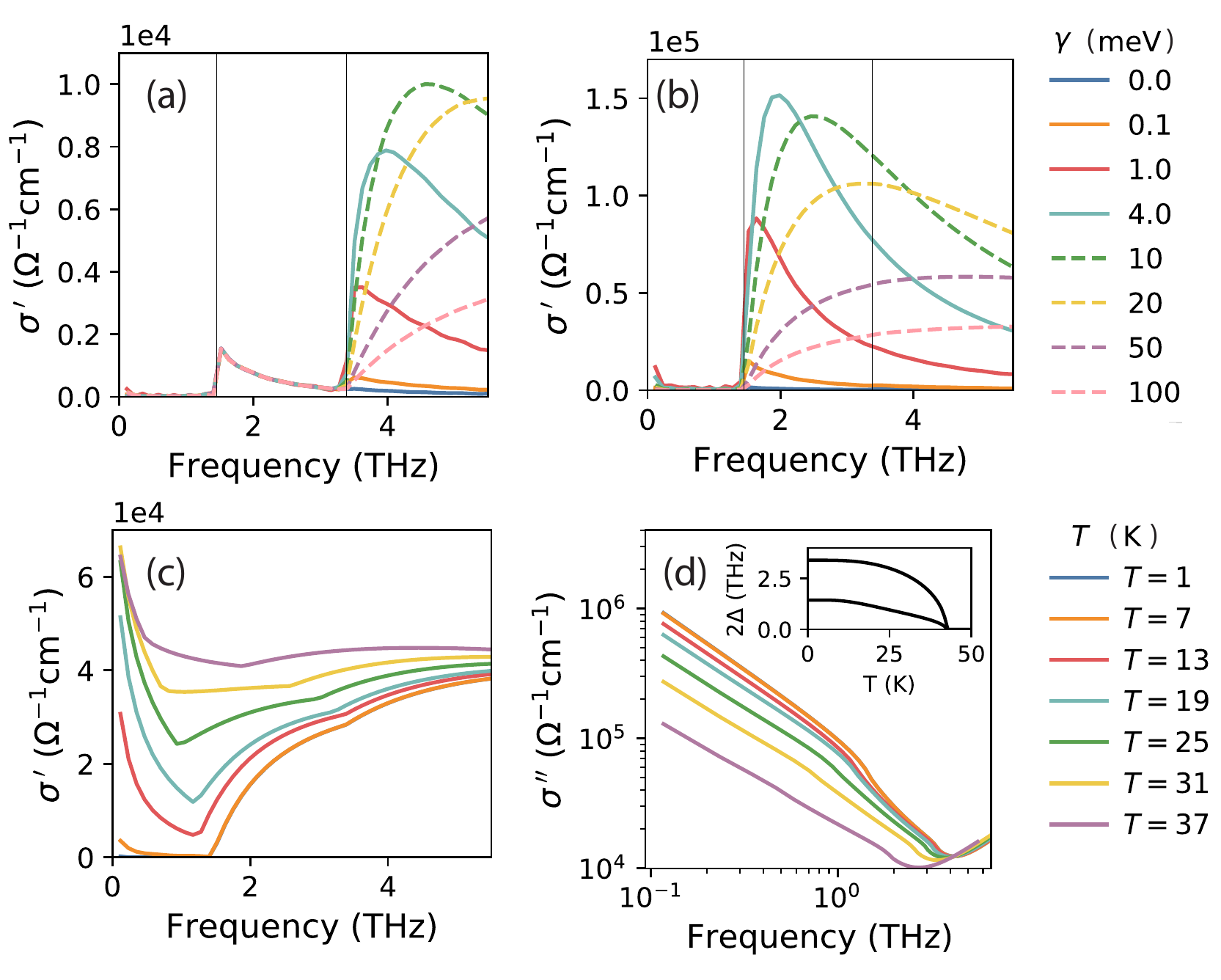}
	\caption{(a),(b) Real part $\sigma'$ 
		of linear response optical conductivity of a two-band superconductor 
	for various impurity scattering rates at $T=\SI{4}{\kelvin}$. In panel (a) the impurity of
	concentration of the first band is $\gamma_\pi=\SI{0.01}{\milli\electronvolt}$ and the second-band 
	impurity scattering rates are given by the legend. In panel (b) the legend specifies $\gamma_{\pi}$
	and $\gamma_\sigma=\SI{0.1}{\milli\electronvolt}$. 
	(c),(d) $\sigma'$ and $\sigma''$ for various temperatures at $\gamma_1=\SI{100}{\milli\electronvolt}$ and $\gamma_2=\SI{50}{\milli\electronvolt}$. The imaginary part follows a $1/\omega$ power-law at small frequencies.}
\label{fig:cond-lin2}
\end{figure}

The linear response optical conductivity of multi-band superconductors is additively composed of contributions 
from the two bands, $\sigma(\omega) = \sigma_\pi + \sigma_\sigma$, where the band-specific conductivities are determined by a straightforward
generalization of Eq.~(\ref{eqn:sigma-expl}). Figures~\ref{fig:cond-lin2}(a-b) show optical conductivities for various 
different combinations of band impurity concentrations.

Experimental measurements of the optical conductivity of MgB$_2$ below $T_C$ show a clear absorption gap below $2\Delta_\pi$ and 
a dome shaped onset above $2\Delta_\pi$. A second onset at $\omega= 2\Delta_\sigma$ has 
so far not been observed. Our simulations reproduce these findings in two different parameter regimes: in the dirty-clean limit, 
where only the first gap contributes to $\sigma(\omega)$, and in the dirty-dirty limit shown in Fig.~\ref{fig:cond-lin2}(c-d). 
Latter case only shows a weak onset of the $\sigma$-gap which may be unnoticeable with experimental uncertainties. 
The reason of the subdominant contribution of the second gap lies in the small Fermi surface of the $\sigma$-band.
Explicitly, this can be seen from the prefactor $v_{F_i} N_i$ in Eq.~(\ref{eqn:sigma-expl}). For our choice of parameters, which
include a high estimate of $\epsilon_{F_\sigma}$, this yields a suppression of 
the $\sigma$-gap conductivity by a factor $v_{F_\pi} N_\pi / v_{F_\sigma} N_\sigma =6.6$. 
For a more conservative estimate of $\epsilon_{F_\sigma}$, the suppression should be even more pronounced.

\subsection{Collective modes}

\begin{figure}[t]
  \centering
  \includegraphics[width=\columnwidth]{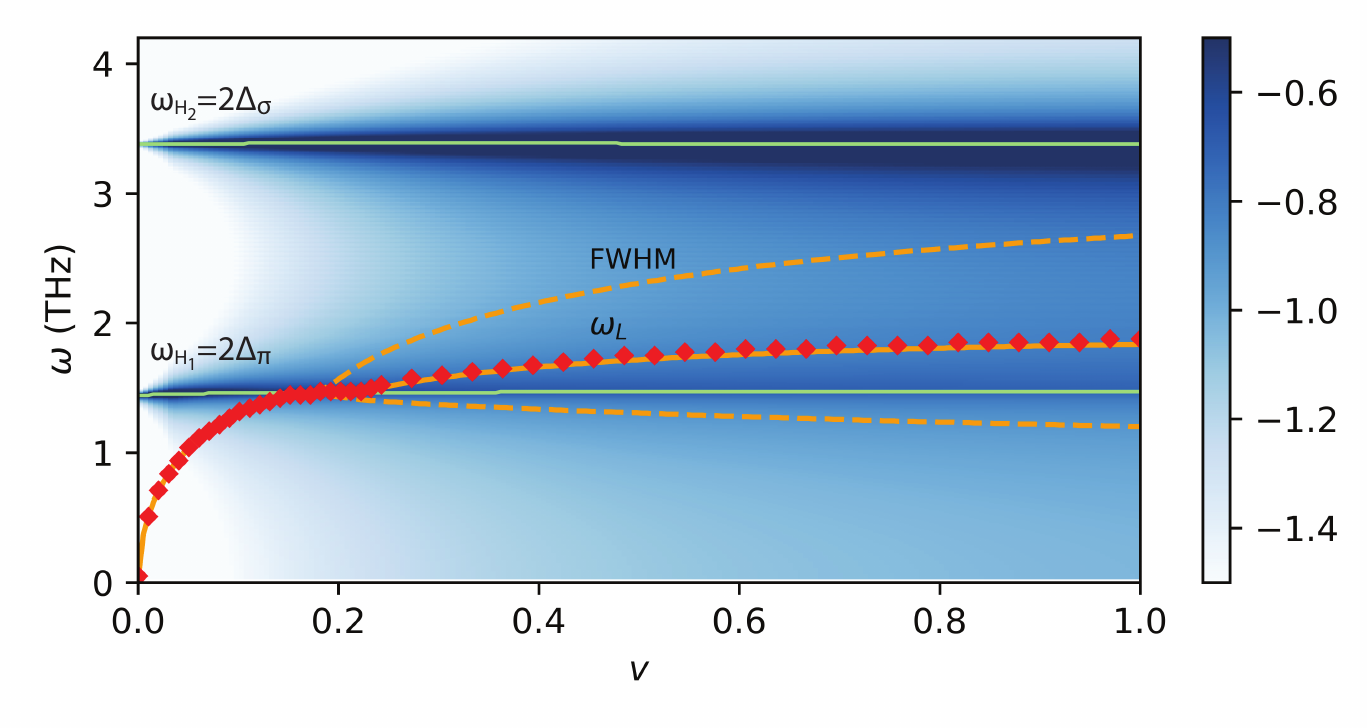}
  \caption{Logarithmic plot of resonance spectrum of Higgs and Leggett modes as a function of interband coupling parameter $v$. False-color
  plot was computed within the effective action formalism. Solid green line shows the frequency of the Higgs resonances. The solid and 
  dashed orange lines mark the maximum and width of the Leggett mode. Red diamonds mark the Leggett oscillation frequencies extracted from a pumped time-dependent density-matrix simulation. The two approaches show excellent agreement. }
  \label{fig:leggett-kernel}
\end{figure}

Pulse induced changes of the two order parameters $\Delta_i$ with $i=\pi,\sigma$ in the two-band case are given by
\begin{align}
  \delta\Delta_i(\omega) &=
  \frac{1}{2}
  \sum_{j\mathbf{kk'}}
  H_{ij}^{-1}(\omega)
	|J_{j\mathbf{kk'}}|^2
	\int d\omega'
	\chi_j^{\sigma_0\sigma_0\sigma_1}(\omega,\omega', \mathbf{k}, \mathbf{k'})
	\nonumber
	\\
	&\qquad \times
	A(\omega') A(-\omega-\omega')\,,
	\label{eq:delta-delta-multiband}
\end{align}
where 
\begin{align}
    	H=\begin{pmatrix}
		\chi_{1}^{\sigma_1\sigma_1}  + 2U_{22} /\det U &
		-2U_{12}/\det U \\
		-2U_{21}/\det U &
	\chi_{2}^{\sigma_1\sigma_1} + 2 U_{11}/\det U &
	\end{pmatrix}
\end{align}
and where susceptibilities $\chi_i^{\sigma_0\sigma_0\sigma_1}, \chi_{i}^{\sigma_1\sigma_1}$ are listed in 
Appendix \ref{apdx:effective-action}.
The gaps exhibit two resonances which are determined by the Higgs propagator. In Fig.~\ref{fig:leggett-kernel}
we show a logarithmic false-color plot of the quantity $|\det H|^{-1}$, responsible for any divergence,
 as a function of frequency $\omega$ and interband coupling strength $v$.
As expected the two resonance energies are at $2\Delta_{\pi}$ and $2\Delta_\sigma$, illustrated
by solid green horizontal lines. Resonances are sharp at 
small $v$ but decrease and broaden in the strong interband coupling regime.

Energy conservation in Eq.~\eqref{eq:delta-delta-multiband} is established by the factor 
$A(\omega') A(-\omega-\omega')$. Oscillation of the gaps is therefore
only possible for a finite overlap of $A^2(\omega)$ with the resonance frequencies. The
matrix structure of $H_{ij}$ further implies that both gaps will oscillate with all excited modes at finite $v$.

\begin{figure*}[t]
  \centering
  \includegraphics[width=0.95\textwidth]{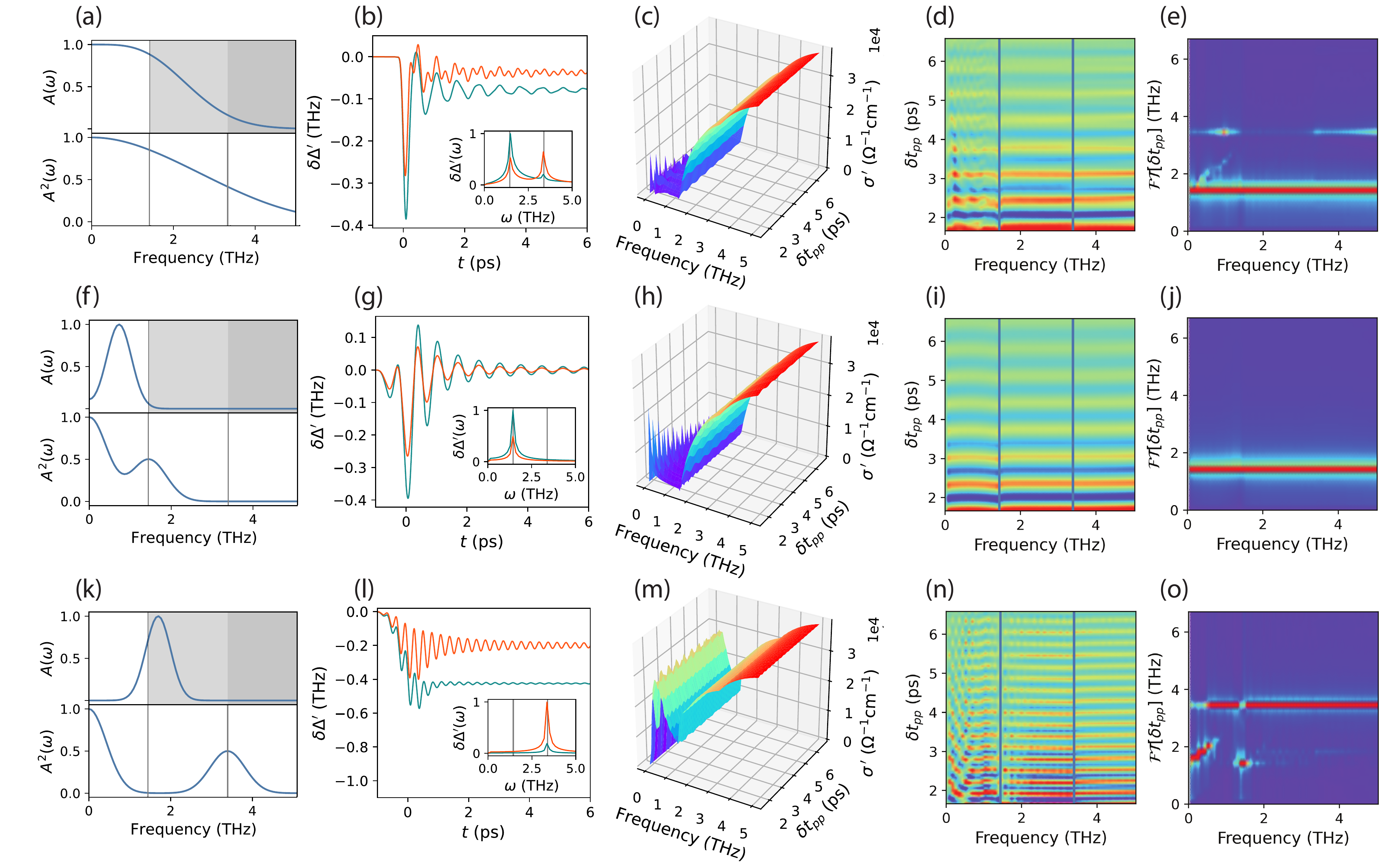}
  \caption{Time-resolved optical conductivity (c),(h),(m) for three optical pulses that resonantly excite 
  (a) both Higgs modes, (f) the lower $\pi$-band Higgs resonance, and (k) the $\sigma$-Higgs mode for an
  interband coupling strength $v=0.2$ in the dirty-dirty limit with $\gamma_\pi=\SI{100}{\milli\electronvolt}$ and 
$\gamma_\sigma=\SI{50}{\milli\electronvolt}$. (b),(g),(l) show the gap oscillations $\delta\Delta'(t)$ as a response to
the pump pulse only. Last two columns show the background subtracted and normalized optical conductivity and the Fourier transform thereof.}
  \label{fig:mgb2-pp}
\end{figure*}

Dynamics of the phase modes $\theta_i$ in the frequency domain are determined by
\begin{equation}
    \delta\theta_i(\omega) = \frac{1}{2} \sum_{j} \frac{s_j e^2}{2 m_j \omega^2}
    L^{-1}_{ij}(\omega) \chi_j^{\sigma_3\sigma_3}(\omega) A^2(\omega) \,.
    \label{eq:phasedynamics}
\end{equation}
Due to the Anderson-Higgs mechanism only the dynamics of the phase difference
$\delta \varphi = \delta\theta_\pi -  \delta\theta_\sigma$ is physical. 
Inserting Eq.~\eqref{eq:phasedynamics} yields the expression
\begin{align}
    \delta \varphi(\omega) &=  
    \frac{1}{4}
    A^2(\omega)
    \left(
    \frac{s_\pi}{m_\pi}-\frac{s_\sigma}{m_\sigma}
    \right)
    \nonumber
    \\
    &\quad
    \times
    \left[
    \omega^2 + \frac{8\Delta_\pi\Delta_\sigma v }{ U_{\sigma\sigma}-v^2 U_{\pi\pi}} \frac{\chi_\pi^{\sigma_3\sigma_3}+\chi_\sigma^{\sigma_3\sigma_3}}{\chi_\pi^{\sigma_3\sigma_3}\chi_\sigma^{\sigma_3\sigma_3}}
    \right]^{-1}\,.
\end{align}
Solid and dashed orange lines in Fig.~\ref{fig:leggett-kernel} trace the maximum and full width at half max (FWHM)
of $\delta \varphi(\omega)/  A^2(\omega)$. 
 Red diamonds are the dominant oscillation frequency of the phase
\begin{equation}
    \delta \varphi(t) \approx \frac{\delta \Delta''_\pi}{\Delta_\pi} - \frac{\delta \Delta''_\sigma}{\Delta_\sigma}
\end{equation}
evaluated by computing $\delta\Delta''_i$ in a time-dependent density matrix formulation for a broadband optical pulse.
The two methods show excellent agreement.
At small coupling the phase exhibits completely undamped oscillations
due to the absence of decay channels.
The Leggett frequency $\omega_L$ increases for stronger coupling. Once its energy reaches the quasiparticle threshold it is increasingly damped and the resonance broadens.

The present results reproduce the findings of Refs. \cite{Murotani2017, 2016Benfatto-Leggett} which were obtained 
in the clean limit. This should come at no surprise since impurities do not change the frequency of the collective resonance within 
the MB approach and additionally the Leggett mode only couples diamagnetically to electromagnetic fields.

\subsection{Pump-probe simulations}


We proceed to model the pump-probe response of a two-band superconductor. Analogous to the single-band
case we consider non-linear contributions to the optical conductivity and pump the system with an intense pulse. 
After some time delay $\delta t_{pp}$, the optical conductivity is probed in the linear response regime by a weak probe pulse.

In Fig.~\ref{fig:mgb2-pp} we adopt the dirty-dirty limit with $\gamma_\pi=\SI{100}{\milli\electronvolt}$ and 
$\gamma_\sigma=\SI{50}{\milli\electronvolt}$
as a potential description of MgB$_2$ and select various pump pulses shown in the leftmost panels.
Gray and dark gray areas illustrate the onset
of the quasiparticle continuum of the two bands. Lower panels show $A^2(\omega)$ where
Higgs resonance frequencies are marked by gray vertical lines.
The second column shows the gap dynamics $\delta\Delta_i(t)$ following the pump pulse. The third
column shows the real part of the time-resolved non-linear optical conductivity $\sigma'(\omega,\delta t_{pp})$.
Last two columns plot isolated and normalized conductivity oscillations, obtained by subtraction of the
constant dome shaped background, as well as their Fourier transform along the $\delta t_{pp}$ axis.

The first pump has a broad frequency spectrum such that it overlaps with both Higgs resonances. 
Following the excitation, both gaps oscillate with both frequencies. The overlap of $A(\omega)$ with the quasiparticle continuum
induces a small drop of $\delta \Delta'$. The optical conductivity shows oscillations in the pump-probe delay $\delta t_{pp}$ with mostly $2\Delta_\pi$ and a small $2\Delta_\sigma$ component. We attribute 
the subdominance of the $\delta\Delta_\sigma$-contribution to the small $\sigma$-band Fermi surface.

For a narrowband pulse centered at $\omega=\Delta_\pi$ (second row), we observe $2\Delta_\pi$ oscillations only. 
Here, the pulse $A(\omega)$ does not overlap with the quasiparticle continuum. As a result, the 
gap oscillates around its equilibrium value $\Delta_\infty=\Delta$.

When the narrowband pulse is centered around the second Higgs resonance at $\omega=2\Delta_\sigma$ (third row), 
the gap performs $2\Delta_\sigma$ oscillations only. However, the response is weak and numerically hard to resolve
in the optical conductivity.

We note that simulations presented in Fig.~\ref{fig:mgb2-pp} show no signatures of the Leggett mode.
This is because Higgs and quasiparticle contributions dominate the optical response even at small disorder.

\subsection{Third harmonic generation} 
\label{sec:THG}
\begin{figure}[t]
  \centering
  \includegraphics[width=0.9\columnwidth]{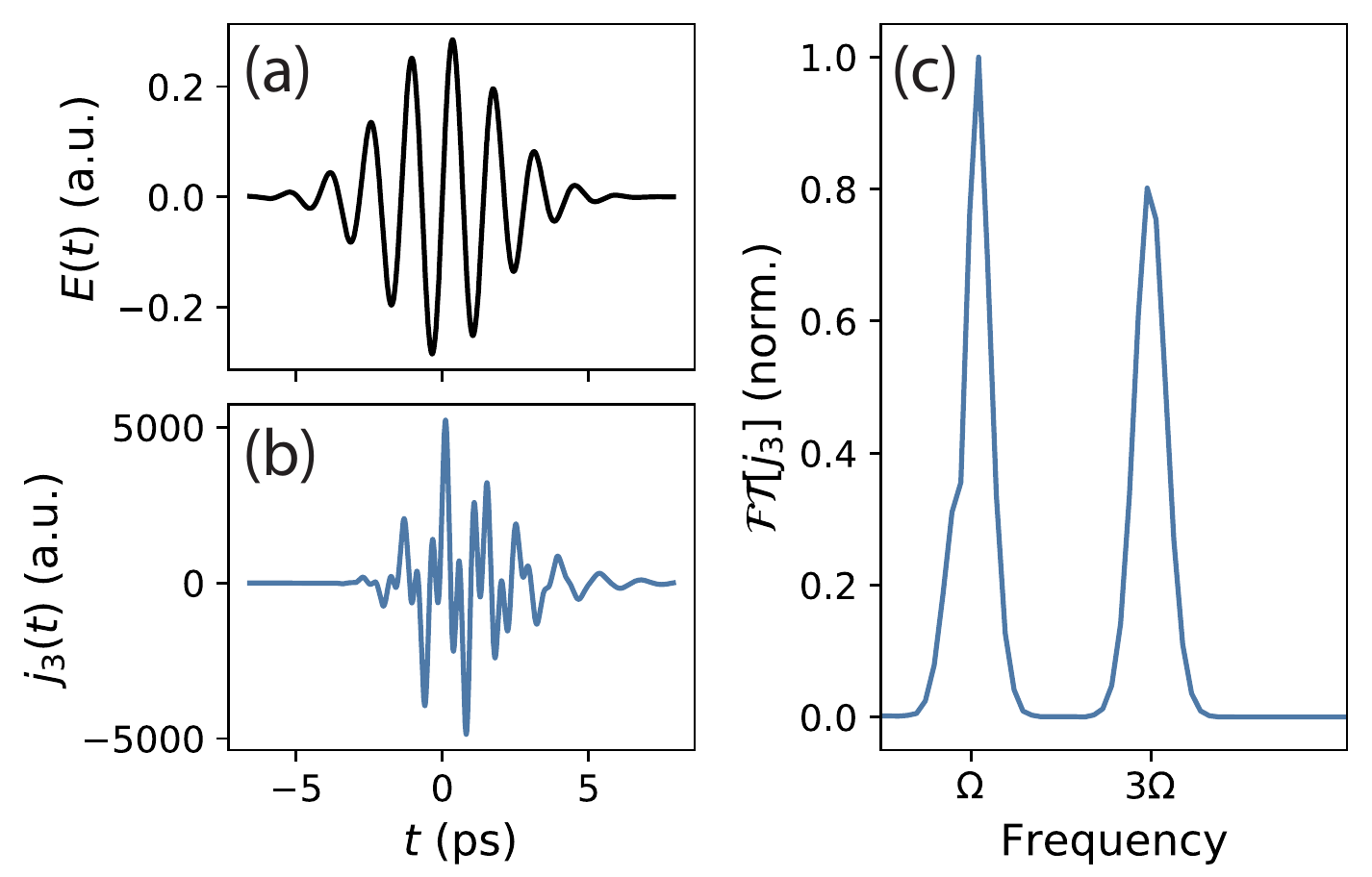}
  \caption{(a) Realistic multicycle pulse of main frequency $\Omega$ fed into time-dependent density matrix simulation. 
  (b) Simulated third order current $j_3(t)$. (c) Next to the original $\Omega$ component,
   the Fourier transform $|j_3(\omega)|$ reveals an additional $3\Omega$ component.}
  \label{fig:thg-expl}
\end{figure}
\begin{figure}[ht]
  \centering
  \includegraphics[width=0.95\columnwidth]{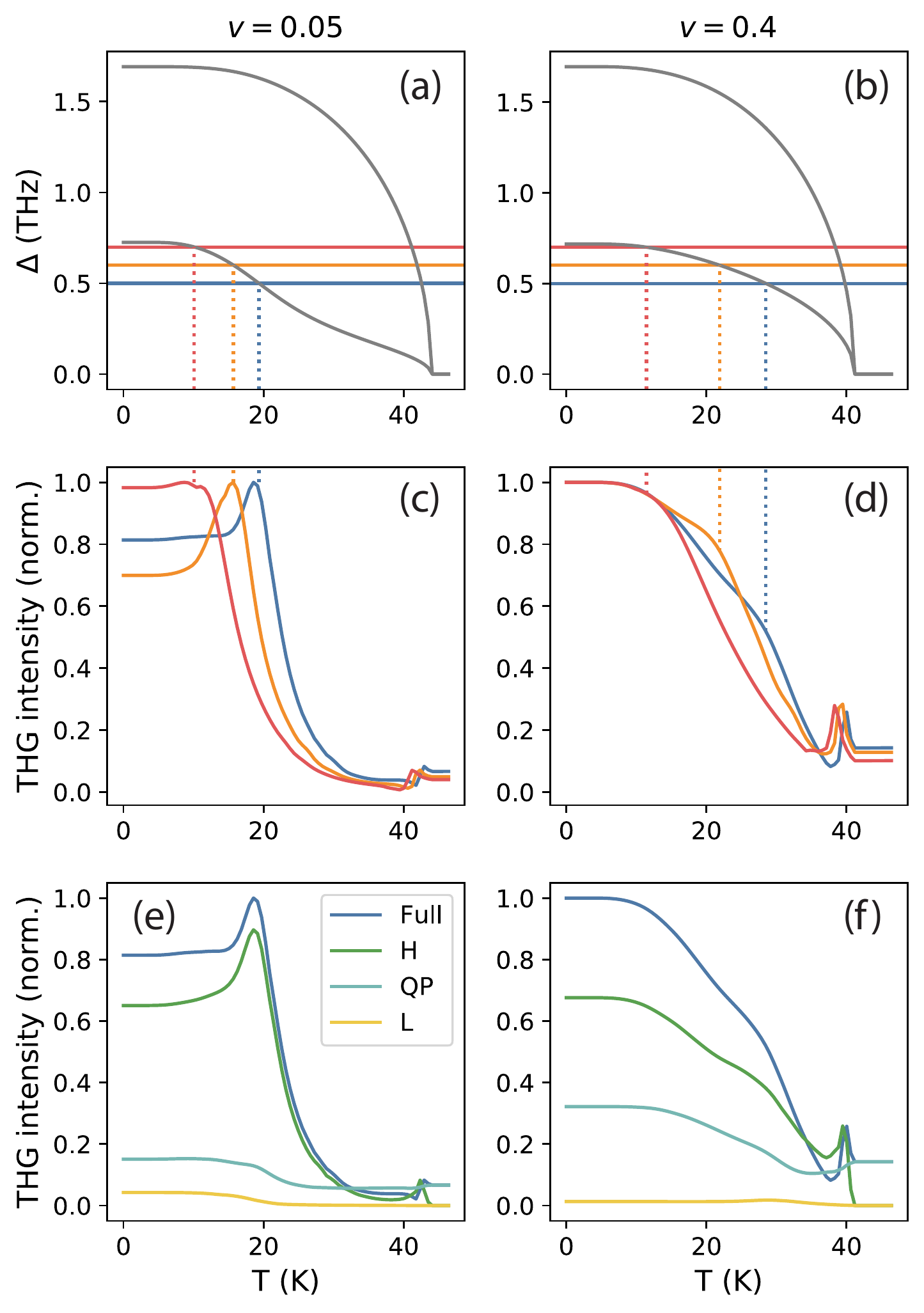}
  \caption{(a),(b) Temperature dependence of the BCS gaps at $v=0.05$ and $v=0.4$. Horizontal lines
  mark the three pulse frequencies $\Omega=0.5,0.6,\SI{0.7}{\tera\hertz}$.
  (c),(d) THG current as a function of temperature for three pulse frequencies $\Omega_j$. We take the THG current as
  $j_3(\omega=3\Omega)$, i.e. the amplitude of the second peak in Fig.~\ref{fig:thg-expl}(c) and sweep temperature.
  (e),(f) Decomposition of the THG signal for pulse of $\Omega=\SI{0.5}{\tera\hertz}$ in Higgs (H), quasiparticle (QP) and Leggett (L) contributions. 
  The main contribution 
  stems from the collective Higgs mode in both the weak coupling (left) and strong coupling case (right).}
  \label{fig:THGpaul}
\end{figure}

Finally, we simulate the non-linear response of a multiband superconductor in a THG setup within the time-dependent density
matrix framework. We model a realistic multi-cycle pulse of frequency $\Omega$, exemplary shown
in Fig.~\ref{fig:thg-expl}, and compute the third order current $j(t)\big|_3$. The Fourier transform of $j(t)\big|_3$ reveals a $3\Omega$ third harmonic (TH) component next to the original first harmonic (FH) peak. 

We adopt the dirty-dirty band description
of MgB$_2$ with $\gamma_\pi=\SI{100}{\milli\electronvolt},\gamma_\sigma=\SI{50}{\milli\electronvolt}$ and choose two different
interband coupling strengths, $v=0.05$ and $v=0.4$. Then, we sweep temperature to investigate the resonant behaviour of the TH component. We consider three pulses of frequencies $\Omega=0.5,0.6,0.7\,\si{\tera\hertz}$ and expect the TH component to be resonantly enhanced when $2\Omega=2\Delta_i$.

Figs.~\ref{fig:THGpaul}(a-b) show the temperature dependence of the BCS gap. Horizontal lines mark pulse frequencies $\Omega$ used in independent simulations. Resonance conditions are satisfied
at intersections with a gap. In the second row, Figs.~\ref{fig:THGpaul}(c-d), the amplitude of the TH peak is found as a function of temperature. In the weak coupling case, $v=0.05$, the THG signal for the lower two frequencies exhibits a pronounced peak at the resonance condition for the lower gap. The THG signal peak of the largest frequency is less pronounced, as this frequency is almost equal to the lower gap for a range of temperatures. We also observe much smaller peaks at temperatures
where pulses are in resonance with the larger $\sigma$-gap.

In the strong coupling case, $v=0.4$, we no longer observe a peak-like resonance for the lower $\pi$-band gap. This can be understood as a result
of broadening of the Higgs resonance at large $v$, further discussed in Appendix~\ref{apdx:thg}. 
The $\sigma$-gap still induces a sharp resonance peak, albeit small in comparison to the low-temperature signal.

Panels (e),(f) of Fig.~\ref{fig:THGpaul} decompose the THG signal for the $\Omega=0.5$ THz pulse into contributions from the Higgs mode, quasiparticles, and Leggett mode. The Leggett mode contribution is found numerically by considering only the diamagnetic component of the current $j_D\big|_3$. The quasiparticle contribution is found by forcing $\delta \Delta_i =0$ when solving the equations of motion, removing the self-consistency condition that induces collective modes. 
In both the weak coupling and large-$v$ case the THG response is dominated by the Higgs mode. The relative contribution of quasiparticles increases in the strong interband coupling
regime. The Leggett contribution is vanishingly small.

The present results are interesting when compared to the experimental findings of Ref.~\cite{Kovalev2020}. 
Our results affirm the claim that the THG response is mainly attributed to the Higgs resonance of the $\pi$-band.
The small contribution of the the $\sigma$-band Higgs mode and the Leggett mode in our simulation 
is consistent with the experiment where no signatures of the Leggett or second Higgs mode were observed.
We have further computed the THG response in the dirty-clean limit where we found nearly identical results, apart from the absence of the
small $\sigma$ resonance peak at temperatures close to $T_C$.

The failure of our theory to produce resonance peaks of the $\sigma$-Higgs mode at large $v$ suggests that the MB approximation might not correctly describe the THG response in the strong coupling limit as assumed for MgB$_2$ \cite{Blumberg2007,2016Benfatto-Leggett}. A recent study has found that incorporating impurities beyond Mattis-Bardeen as random onsite-energies in a lattice model shows a stronger contribution of quasiparticles \cite{Seibold2020}. This, however, is beyond the scope of this paper and will be explored in future investigations.

\section{Conclusion}
\label{sec:conclusion}
We have calculated the time-resolved optical response of dirty multiband superconductors. We have incorporated impurity scattering within the Mattis-Bardeen approximation that effectively broadens the photon momentum distribution of the optical pulse. This approach is known to accurately describe superconductors \cite{PhysRev.154.414,Matsunaga2013}, yet deviations in the strong disorder regime are possible \cite{PhysRevLett.109.107003,PhysRevB.88.180505,Seibold2017}. The response was calculated within two different frameworks. First, the time-evolution of the system after an excitation with a light pulse was calculated explicitly using a time-dependent density matrix formalism. Here, the Mattis-Bardeen ansatz enters through a replacement of the matrix element of the current operator with a Lorentzian-shaped momentum transfer distribution \cite{Murotani2019}. In the second approach, we calculated the relevant susceptibilities in a diagrammatic formalism derived from an effective action approach. Here, impurities 
enter through a paramagnetic electromagnetic coupling vertex that carries external momentum. As a consequence, additional diagrams arise that usually vanish in the clean limit.
While both approaches yield equivalent results, the diagrammatic approach allows to understand the relevant processes in more detail 
and is numerically more efficient in certain cases.

In accordance with previous literature \cite{Cea2016,Murotani2019,Seibold2020}, we find that the collective Higgs response is drastically enhanced
even for small impurity concentrations. The Leggett mode is unaffected by impurity scattering
and hence becomes subdominant. This may change slightly when realistic, non-parabolic bandstructures and weak violations 
of particle-hole symmetry are taken into account. An interesting further question is the inclusion of Coulomb interactions within the MB approach.

As a first result, we calculated the dynamics of superconducting order paramater
of a single-band superconductor in the dirty limit excited by a short THz quench pulse. 
Using a second probe-pulse after a variable time delay, we further computed the time-resolved optical conductivity. 
Both quantities show oscillations with the Higgs frequency $\omega_H = 2\Delta$. The optical response is in good agreement to measurements on NbN \cite{Matsunaga2013}. 

Extending the model to two bands, we studied pump-probe optical conductivities of the model 
as an effective low-energy description of MgB$_2$ for various impurity limits of the $\pi$- and $\sigma$-band. We found that experimental
results are well reproduced either when both bands are dirty, or when only the lower $\pi$-band is dirty. Here, the collective contribution
to the non-linear optical response is always dominated by the amplitude mode of $\Delta_\pi$. 

Finally, we presented the third-harmonic generation (THG) response of the two-band model. Interestingly, results obtained in the
weak interband coupling regime seem to match available experimental data, showing a pronounced THG resonance mostly due to the $\pi$-band Higgs mode \cite{Kovalev2020}.
However, our theory shows deviations in the strong interband coupling case, believed to be representative of MgB$_2$, where no obvious $\pi$-Higgs-resonance is present.

In summary, we have presented simulations of time-resolved optical conductivities and modelled THG resonance experiments 
within the Mattis-Bardeen approximation
using both a diagrammatic approach and a time-dependent density matrix formalism for single-band and two-band superconductors.
As studies of collective excitations in superconductors
with THz spectroscopy become more and more common,
it is important to understand the correct excitation
scheme in the presence of impurities.
The Mattis-Bardeen approximation,
as shown in this work for
either a density-matrix formalism
or a diagrammatic calculation,
allows to describe the effects of impurities
in a simple way.
This will help further studies in achieving more realistic models of experimental results.

\begin{acknowledgments}
We are indebted to M. Daghofer, M. Puviani, A. Schnyder, and R. Shimano for illuminating discussions. We thank the Max Planck-UBC-UTokyo Center for Quantum
Materials for fruitful collaborations and financial support. R.H. acknowledges the Joint-PhD program of the University of British Columbia and the University of Stuttgart. P.F. thanks the UBC Science Co-op program for making an internship possible, and acknowledges computational resources provided by Advanced Research Computing at the University of British Columbia. 
\end{acknowledgments}

\newpage
\appendix
\begin{widetext}

\section{Derivation of the Effective Action}
\label{apdx:effective-action}
The problem is stated with the partition function
\begin{eqnarray}
	\mathcal{Z} = \int\mathcal{D}(c^\dagger c) e^{-S} \quad \text{with} \quad S = \int_{0}^{\beta}d\tau
	\left( 
		\sum_{i\mathbf{k}\sigma}^{} 
		c_{i\mathbf{k}\sigma}^\dagger  \partial_\tau
		c_{i\mathbf{k}\sigma} 
		+ \mathcal{H}
	\right) \,.
\end{eqnarray}
We decouple the interacting term in the pairing channel via the Hubbard
Stratonovich transformation
\begin{align}
	&\exp \left(
	\sum_{ij}^{}\sum_{\mathbf{kk'}}^{}  c_{i\mathbf{k}\uparrow
	}^\dagger
	c_{i-\mathbf{k}\downarrow }^\dagger
	U_{ij}
	c_{j-\mathbf{k'}\downarrow
	}c_{j\mathbf{k}'\uparrow }
	\right)
	\nonumber
	\\
	&\qquad\qquad=
	\int_{}^{}\mathcal{D}(\bar{\Delta}_i \Delta_i)
	\exp \left( -\sum_{i\mathbf{k}}
		\left[
			\sum_{j}^{}
		\bar{\Delta}_i U^{-1}_{ij} \Delta_j
		-\left(
			\bar{\Delta}_i c_{i\mathbf{k}\uparrow
			}c_{i-\mathbf{k}\downarrow }  +
			\Delta_i c_{i\mathbf{-k}\downarrow  }^\dagger c_{i\mathbf{k}\uparrow }^\dagger
	\right)
		\right]
\right)\,.
\end{align}
The bosonic field $\Delta$ is complex, i.e. it permits amplitude and phase
fluctuations. We decompose it into real fields and additional express fluctuation with respect to the meanfield saddlepoint,
$\Delta_i(\tau) \rightarrow (\Delta_i^{eq}+\Delta_i(\tau))e^{i\theta_i(\tau)}$, $\bar{\Delta}_i(\tau) \rightarrow (\Delta_i^{eq}+\Delta_i(\tau))e^{-i\theta_i(\tau)}$. Note that we are neglecting spatial
fluctuations of the collective fields, i.e. $\Delta_i(\tau)$, $\theta_i(\tau)$ depend on time
only. The action is
\begin{eqnarray}
	S&=&\sum_{ij}^{}U_{ij}^{-1} \int d\tau \, 
	\left(
	\Delta_i^{eq} \Delta_j^{eq}
	+
	\Delta_i(\tau) \Delta_j(\tau) 
	\right)
	e^{-i(\theta_1(\tau)-\theta_2(\tau))} 
	\nonumber
	\\
	&+&
	    \sum_{i\mathbf{k}\sigma}^{} 
        \int_{0}^{\beta}d\tau
		\left(
		c_{i\mathbf{k}\sigma}^\dagger  
		\left[
		\partial_\tau 
		+ \epsilon_{i\mathbf{k}}
		\right]
		c_{i\mathbf{k}\sigma} 
		-
			\Delta_i(\tau)e^{-i\theta_i(\tau)} c_{i\mathbf{k}\uparrow
			}c_{i-\mathbf{k}\downarrow }  -
			\Delta_i(\tau)e^{i\theta_i(\tau)} c_{i\mathbf{-k}\downarrow  }^\dagger c_{i\mathbf{k}\uparrow }^\dagger  
	\right)
		+ 
		\int_{0}^{\beta}d\tau \, \mathcal{H}_1\,.
\end{eqnarray}
We express it in Nambu
basis $\Psi_{i\mathbf{k}}^\dagger(\omega_n) =\left( c_{i\mathbf{k}\uparrow
}^\dagger \,, c_{i-\mathbf{k}\downarrow } \right)$,
\begin{eqnarray}
	S=\sum_{ij}^{}U_{ij}^{-1}
	\int d\tau \,
	\left(
	\Delta_i^{eq} \Delta_j^{eq}
	+
	\Delta_i(\tau) \Delta_j(\tau) 
	\right)
	e^{-i(\theta_1-\theta_2)}
	-
	\sum_{i\mathbf{kk'} }\int d\tau 
	\Psi^\dagger_{i\mathbf{k}}(\tau) 
	G^{-1}_{i}(\mathbf{k}\mathbf{k'},\tau)
	\Psi_{i\mathbf{k'}}(\tau)\,,
\end{eqnarray}
where
\begin{align}
G_{0,i}^{-1}
	=
	\delta_{\mathbf{kk'}}
	\begin{pmatrix}
		\partial_\tau -\xi_{i\mathbf{k}} & (\Delta_i^{eq}+\Delta_i(\tau))e^{i\theta(\tau)}  \\ 
		(\Delta_i^{eq}+\Delta_i(\tau)) e^{-i\theta(\tau)} & \partial_\tau +\xi_{i\mathbf{k}}
			\end{pmatrix}
			+J_{i\mathbf{kk'}} \cdot \mathbf{e} \, A(\tau) \sigma_0
		- \frac{s_i e^2}{2m_i} A^2(\tau) \delta_{\mathbf{kk'}} \sigma_3 \,.
\end{align}
Integrating out the Fermions gives
\begin{eqnarray}
		S=\sum_{ij}^{}U_{ij}^{-1}
	\int d\tau \,
	\left(
	\Delta_i^{eq} \Delta_j^{eq}
	+
	\Delta_i(\tau) \Delta_j(\tau) 
	\right)
	e^{-i(\theta_1-\theta_2)}
	- \sum_{i}^{} \text{Tr} \ln \left[
	-G_i^{-1} \right] \,,
	\label{eq:fullaction}
\end{eqnarray}
where the trace is performed over time, momenta, and Nambu indices, but not over band-indices $i$.
To separate amplitude $\Delta(\tau)$ and phase $\theta(\tau)$ fields, we introduce a unitary transformation
$V=\exp (i\theta(\tau) \sigma_3/2)$ \cite{Thouless.1995},
\begin{align}
    \text{Tr} \ln \left[- G^{-1}_i\right] 
    = \text{Tr} \ln\left[- G^{-1}_i V V^\dagger \right]
    = \text{Tr} \ln \left[-V^\dagger G^{-1}_i V \right]
    =    \text{Tr} \ln \left[-\tilde{G}^{-1}_i\right] \,.
\end{align}
We split $\tilde{G}^{-1}_i$ into a meanfield part, $G_{0,i}^{-1}$, and all remaining contributions $\Sigma_i$. In frequency space, this gives
\begin{eqnarray}
	\tilde{G}^{-1}_i &=& G_{0,i}^{-1} - \Sigma_i\,,
	\\
	G_{0,i}^{-1}(\mathbf{k}\omega_n,\mathbf{k'}\omega_m) 
	&=&
	\left[ 
		i\omega_n -
		\xi_{i\mathbf{k}} \sigma_3 + \Delta_i^{eq} \sigma_1 
	\right] 
	\delta_{\mathbf{kk'}}\delta_{\omega_n,\omega_m}\,,
	\\
	\Sigma_i(\mathbf{k}\omega_n,\mathbf{k'}\omega_m) &=& 
	-\Delta_i(\omega_n-\omega_m) \sigma_1 \delta_{\mathbf{kk'}}  - \frac{i}{2} \theta_i(\omega_n-\omega_m)
		(i\omega_n-i\omega_m) \sigma_3 \delta_{\mathbf{kk'}}
	\nonumber
	\\
	&&
		-J_{i\mathbf{kk'}} \cdot \mathbf{e} \, A(\omega_n-\omega_m) \sigma_0
		+ \frac{s_i e^2}{2m_i} A^2(\omega_n-\omega_m) \delta_{\mathbf{kk'}} \sigma_3 \,.
\end{eqnarray}
Note that phase fluctuations $\theta$ live in the $\sigma_3$ channel, 
i.e. the charge channel.

We expand $S$ at Gaussian level. To compute currents
$j\big|_1$ and $j\big|_3$ we additionally keep terms up to fourth
order in the classical field $A$. In expanding the trace, we use
\begin{eqnarray}
	\text{Tr}\ln\left( -\tilde{G}^{-1} \right) 
	=
	\text{Tr}\ln\left( -G_0^{-1}(1 - G_0\Sigma) \right) 
	=
	\text{Tr}\ln\left( -G_0^{-1} \right) 
	-\text{Tr}
	\sum_{n=1}^{\infty}
	\frac{1}{n}
	\left( 
	G_0 \Sigma
	\right)^n \,.
\end{eqnarray}
The quadratic action is given by the terms

\begin{equation}
    S[\Delta_i,\theta_i, A] = S_{MF} + S_\Delta + S_{\Delta,A} + S_\theta + S_{\theta,A} + S_{\text{QP,dia}} + S_{\text{QP,para}}\,,
    \label{eq:seffappdx}
\end{equation}
which are explicitly
\begin{eqnarray}
    S_{MF}&=&\sum_{ij}^{}U_{ij}^{-1} \Delta_i^{eq} \Delta_j^{eq} - \sum_{i}^{} \text{Tr} \ln \left[
	-G_{0,i}^{-1} \right]\,,
	\\
    \label{eqn:hl1}
	S_{\Delta} &=& 
	\frac{1}{2}
	\sum_{ij \Omega_m}^{}
		\Delta_i(-\Omega_m)
	\begin{pmatrix}
		\chi_{1}^{\sigma_1\sigma_1}(\Omega_m)  + 2U_{22} /\det U &
		-2U_{12}/\det U \\
		-2U_{21}/\det U &
	\chi_{2}^{\sigma_1\sigma_1}(\Omega_m) + 2 U_{11}/\det U &
	\end{pmatrix}_{ij}
		\Delta_j(\Omega_m)
		\nonumber
	\\
	&=& 
	\frac{1}{2}
	\sum_{ij \Omega_m}^{}
		\Delta_i(-\Omega_m)
		H_{ij}^{-1}(\Omega_m)
		\Delta_j(\Omega_m)\,,
		\label{eq:higgsmatrix}
	\\
	S_{\Delta,A} &=&
	-\sum_{i \mathbf{kk'}}\sum_{\Omega_m\Omega_l} 
	|J_{i\mathbf{kk'}}|^2
	\chi_i^{\sigma_0\sigma_0\sigma_1}(\Omega_m,\Omega_l, \mathbf{k}, \mathbf{k'})
	A(\Omega_l) A(-\Omega_m-\Omega_l) \Delta(\Omega_m)\,,
	\\
	S_{\theta} &=& 
	-\frac{1}{2}
	\sum_{ij \Omega_m}^{}
	\frac{-i\Omega_m}{2}
		\theta_i(-\Omega_m)
	\begin{pmatrix}
		-\chi_{1}^{\sigma_3\sigma_3}(\Omega_m) + \frac{\lambda}{\Omega_m^2} &
		-\frac{\lambda}{\Omega_m^2} \\
		-\frac{\lambda}{\Omega_m^2} &
		-\chi_{2}^{\sigma_3\sigma_3}(\Omega_m) + \frac{\lambda}{\Omega_m^2} &
	\end{pmatrix}_{ij}
	\frac{i\Omega_m}{2}
		\theta_j(\Omega_m)
	\nonumber
		\\
		&=&
			-\frac{1}{2}
	\sum_{ij \Omega_m}^{}
	\frac{-i\Omega_m}{2}
		\theta_i(-\Omega_m)
		\left[
	\begin{pmatrix}
		-\chi_{1}^{\sigma_3\sigma_3}(\Omega_m) &
	 \\
	 &
		-\chi_{2}^{\sigma_3\sigma_3}(\Omega_m) &
	\end{pmatrix}_{ij}
	+ J_{ij}
	\right]
	\frac{i\Omega_m}{2}
		\theta_j(\Omega_m)
	\nonumber
		\\
		&=&
			\frac{1}{2}
	\sum_{ij \Omega_m}^{}
	\frac{-i\Omega_m}{2}
		\theta_i(-\Omega_m)
		L_{ij}^{-1}(\Omega_m)
		\frac{i\Omega_m}{2}
		\theta_j(\Omega_m)\,,
		\label{eq:leggettmatrix}
	\\
	\label{eqn:hl2}
	S_{\theta,A} &=&
	\sum_{i\Omega_m}^{}
	i \frac{s_i e^2}{2m_i} \mathbf{A}^2(\Omega_m)
	\chi^{\sigma_3\sigma_3}_i(\Omega_m) \frac{-i\Omega_m}{2} \theta_i (-\Omega_m)\,,
	\\
    \label{eqn:qp1}
	S_{\text{QP,dia}}^{(2)} &=& 
	-
	\sum_{i}
	\frac{s_i e^2}{2m_i} \chi_i^{\sigma_3} A^2(0)\,,
	\\ 
	S_{\text{QP,dia}}^{(4)}
	&=&
	 \frac{1}{2}\sum_{i \Omega_m}^{}
	\left( 
	\frac{s_i e^2}{2m_i} 
	\right)^2
	\chi_i^{\sigma_3\sigma_3}(\Omega_m) A^2(\Omega_m)A^2(-\Omega_m)\,,
	\\
	S_{\text{QP,para}}^{(2)} &=& \frac{1}{2}\sum_{i\mathbf{kk'}\Omega_m}^{}
	|J_{i\mathbf{kk'}}|^2
	\chi_i^{\sigma_0\sigma_0}(\mathbf{k},\mathbf{k'},\Omega_m) A(\Omega_m)
	A(-\Omega_m)\,,
	\label{eq:j2-qp-para}
	\\ 
	\label{eqn:qp2}
	S_{\text{QP,para}}^{(4)}
	&=&
	\frac{1}{4}\sum_{i\mathbf{kk'k''}} \sum_{\Omega_m\Omega_l\Omega_p}
	|J_{i\mathbf{kk'}}|^2
	|J_{i\mathbf{kk''}}|^2
	\xi_i(\mathbf{k},\mathbf{k'},\mathbf{k''},\Omega_m,\Omega_l,\Omega_p) 
	A(\Omega_m)
	A(\Omega_l)
	A(\Omega_p)
	A(-\Omega_m-\Omega_l-\Omega_p)\,.
	\label{eqn:s4qp-para}
\end{eqnarray}
Here, $\lambda = \frac{8\Delta_1\Delta_2 v }{ U_{22}-v^2 U_{11}}$.
The Higgs and Leggett terms, Eqs. (\ref{eqn:hl1}-\ref{eqn:hl2}), are diagrammatically 
shown in Fig.~\ref{fig:diagrams-higgs-leggett}. 
The quasiparticle terms, Eqs. (\ref{eqn:qp1}-\ref{eqn:qp2}) are shown in Fig.~\ref{fig:diagrams-j}. 
The susceptibilities, given by the fermionic bubbles, are

\begin{eqnarray}
  \chi_i^{\sigma_k} &=& 
  \sum_\mathbf{k}\sum_{\omega_n}
  \text{Tr} \left[ 
  G_{0,i}(\omega_n, \mathbf{k}) \sigma_k
  \right]\,,
  \label{eq:susc-begin}
  \\
  \chi_i^{\sigma_k\sigma_l}(\Omega_m, \mathbf{k}, \mathbf{k'}) &=& 
  \sum_{\omega_n}
  \text{Tr} \left[ 
  G_{0,i}(\omega_n, \mathbf{k}) \sigma_k G_{0,i}(\omega_n + \Omega_m, \mathbf{k'}) \sigma_l
  \right]\,,
  \label{eq:x_kl_of_k}
  \\
  \chi_i^{\sigma_k\sigma_l}(\Omega_m) &=& 
  \sum_\mathbf{k}
  \chi_i^{\sigma_k\sigma_l}(\Omega_m, \mathbf{k}, \mathbf{k})\,,
  \label{eq:x_kl}
   \\
  \chi_i^{\sigma_0\sigma_0\sigma_1}(\Omega_m,\Omega_l, \mathbf{k}, \mathbf{k'}) &=& 
  \sum_{\omega_n}
  \text{Tr} \left[ 
  G_{0,i}(\omega_n+\Omega_m, \mathbf{k})
  G_{0,i}(\omega_n+\Omega_m+\Omega_l, \mathbf{k'})
  G_{0,i}(\omega_n, \mathbf{k})
  \sigma_1
  \right]\,,
  \label{eq:x001}
  \\
  \xi_i(\mathbf{k},\mathbf{k'},\mathbf{k''},\Omega_m,\Omega_l,\Omega_p) &=& 
  \sum_{\omega_n}
  \text{Tr} 
  [
        G_{0,i}(\omega_n,\mathbf{k})
        G_{0,i}(\omega_n+\Omega_m,\mathbf{k'})
        \\
        && \times \,
        G_{0,i}(\omega_n+\Omega_m+\Omega_l,\mathbf{k})
        G_{0,i}(\omega_n+\Omega_m+\Omega_l+\Omega_p,\mathbf{k''})
  ]\,,
  \label{eq:susc-end}
\end{eqnarray}
Additional diagrams,
that vanish due to particle-hole symmetry, are listed in Fig.~\ref{fig:diagrams-vanishing}.
\begin{figure}[t]
    \centering
    \includegraphics[width=\columnwidth]{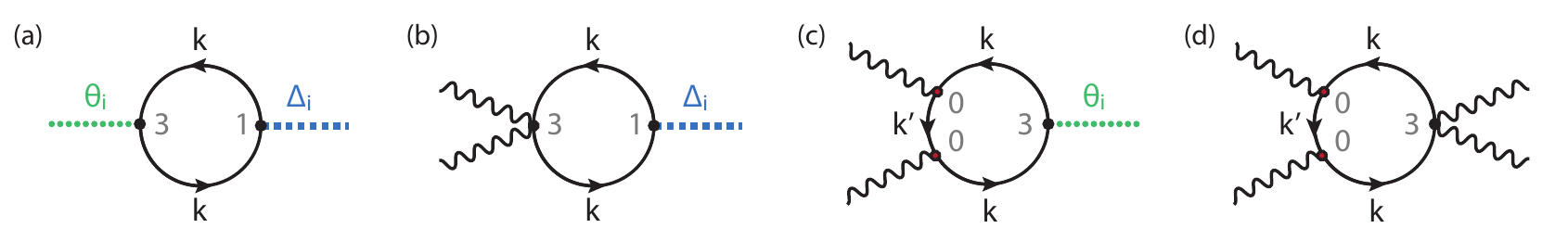}
    \caption{Additional diagrams following from Eq.~(\ref{eq:fullaction}) that vanish in the presence of particle hole symmetry and
    a parabolic dispersion.}
    \label{fig:diagrams-vanishing}
\end{figure}
We proceed to integrate out all collective fields. This gives
\begin{equation}
    S[A] = S_{MF} + \tilde{S}_\Delta + \tilde{S}_\theta + S_{\text{QP,dia}} + S_{\text{QP,para}}
    \label{eq:effaction}
\end{equation}
with
\begin{eqnarray}
  \tilde{S}_\Delta &=& 
  -\frac{1}{2}
  \sum_{ij\Omega_m}
  \left(
 \sum_{\mathbf{kk'} \Omega_l} 
	|J_{i\mathbf{kk'}}|^2
	\chi_i^{\sigma_0\sigma_0\sigma_1}(-\Omega_m,\Omega_l, \mathbf{k}, \mathbf{k'})
	A(\Omega_l) A(\Omega_m-\Omega_l)
  \right) H_{ij}(\Omega_m)
  \nonumber
  \\
  && \times
  \left(
    \sum_{\mathbf{kk'} \Omega_l} 
	|J_{j\mathbf{kk'}}|^2
	\chi_j^{\sigma_0\sigma_0\sigma_1}(\Omega_m,\Omega_l, \mathbf{k}, \mathbf{k'})
	A(\Omega_l) A(-\Omega_m-\Omega_l)
  \right) \,,
  \\
  \tilde{S}_\theta &=& \frac{1}{2} \sum_{ij \Omega_m} \frac{s_i e^2}{2 m_i}\frac{s_j e^2}{2 m_j} A^2(-\Omega_m) A^2(\Omega_m)
  \chi_i^{\sigma_3\sigma_3}(-\Omega_m) L_{ij}(\Omega_m) \chi_j^{\sigma_3\sigma_3}(\Omega_m)
  \,.
\end{eqnarray}
We note that phase and fourth order diamagnetic quasiparticle term combine to give the Leggett contribution
\begin{equation}
    \tilde{S}_L = S_{QP,\text{dia}}^{(4)}+\tilde{S}_\theta = -\frac{1}{2}\sum_{\Omega_m} 
    \frac{\lambda}{4}  
    \left(\frac{s_1}{m_1}-\frac{s_2}{m_2} \right)^2
    \left[ \Omega_n^2 - \lambda \frac{\chi_1^{\sigma_3\sigma_3}+\chi_2^{\sigma_3\sigma_3}}{\chi_1^{\sigma_3\sigma_3}\chi_2^{\sigma_3\sigma_3}} 
    \right]^{-1} A^2(-\Omega_m) A^2(\Omega_m) \,.
\end{equation}
Above equations, obtained by Gaussian integration, have the diagrammatic representation of an RPA summation
shown in Fig.~\ref{fig:diagrams-RPA}. For the Higgs propagator this can be seen by expanding
\begin{equation}
    H = \left[2 U^{-1} + X\right]^{-1} 
    =
    \frac{U}{2}\sum_{n=0}^\infty \left( - X \frac{U}{2} \right)^n \,,
\end{equation}
where $X_{ij}= \chi_i^{\sigma_1\sigma_1} \delta_{ij}$ corresponds to fermionic bubbles and $U_{ij}/2$ corresponds to the to dashed lines. The case of the Leggett mode is analogous.
The currents can now, after analytic continuation of all external frequencies, be computed by a functional derivative of the action,
\begin{eqnarray}
	j(t) = \frac{\delta }{\delta A(t)} \ln Z[A] = -\frac{\delta }{\delta A(t)} S[A] \,.
\end{eqnarray}
In Fourier space this results in
\begin{equation}
    j(-\omega) = - \frac{\delta S[A]}{\delta A(\omega)} \,.
\end{equation}
Latter equality follows in full generality from the chain rule of functional derivation,
\begin{eqnarray}
	\frac{\delta S[A]}{\delta \tilde{A}(\omega)}
	=
	\frac{\delta }{\delta \tilde{A}(\omega)}S[\mathcal{FT}[\tilde{A}]]
	=
	\int_{}^{}dt \underbrace{\frac{\delta S[A]}{\delta A(t)}}_{-j(t)} 
	\underbrace{\frac{\delta \mathcal{FT}^{-1}[\tilde{A}](t)}{\delta
	\tilde{A}(\omega)}}_{e^{-i\omega t}} = 
	-j(-\omega) \,,
\end{eqnarray}
where we have denoted the Fourier transform $\tilde{A}(\omega) = \mathcal{FT}[A](\omega)$ by a tilde.

The Mattis-Bardeen approximation enters by replacing 
\begin{equation}
\sum_{\mathbf{kk'}} \left| J_{i\mathbf{kk'}} \right|
= N_i(0)^2 \int d\epsilon_{\mathbf{k}} d\epsilon_{\mathbf{k'}}
\int \frac{d\Omega_{\mathbf{k}}}{4\pi} \frac{d\Omega_{\mathbf{k'}}}{4\pi}
\left| J_{i\mathbf{kk'}} \right|
\approx
N_i(0)^2 \int d\epsilon_{\mathbf{k}} d\epsilon_{\mathbf{k'}} \frac{(e v_{F_i})^2}{3 N_i(0)} 
	W(\epsilon_{i\mathbf{k}},\epsilon_{i\mathbf{k}'}) 
\end{equation}
according to Eq.~\eqref{eq:MB}.
For the fourth-order paramagnetic quasiparticle contribution, Eq.~(\ref{eqn:s4qp-para}), we follow Ref.~\cite{Murotani2019} 
and further approximate
\begin{equation}
    \sum_{\mathbf{kk'k''}} \left| J_{i\mathbf{kk'}} \right| \left| J_{i\mathbf{kk''}} \right|
    \approx
    N_i(0)^3 \int d\epsilon_{\mathbf{k}} d\epsilon_{\mathbf{k'}} d\epsilon_{\mathbf{k''}}
    \left(
        \int 
        \frac{d\Omega_{\mathbf{k}}}{4\pi} 
        \frac{d\Omega_{\mathbf{k'}}}{4\pi}
        \left| J_{i\mathbf{kk'}} \right|
    \right)
       \left(
        \int 
        \frac{d\Omega_{\mathbf{k}}}{4\pi} 
        \frac{d\Omega_{\mathbf{k''}}}{4\pi}
        \left| J_{i\mathbf{kk''}} \right|
    \right) \,.
\end{equation}

\section{First order currents and optical conductivity}

\begin{figure}[t]
    \centering
    \includegraphics[width=0.45\columnwidth]{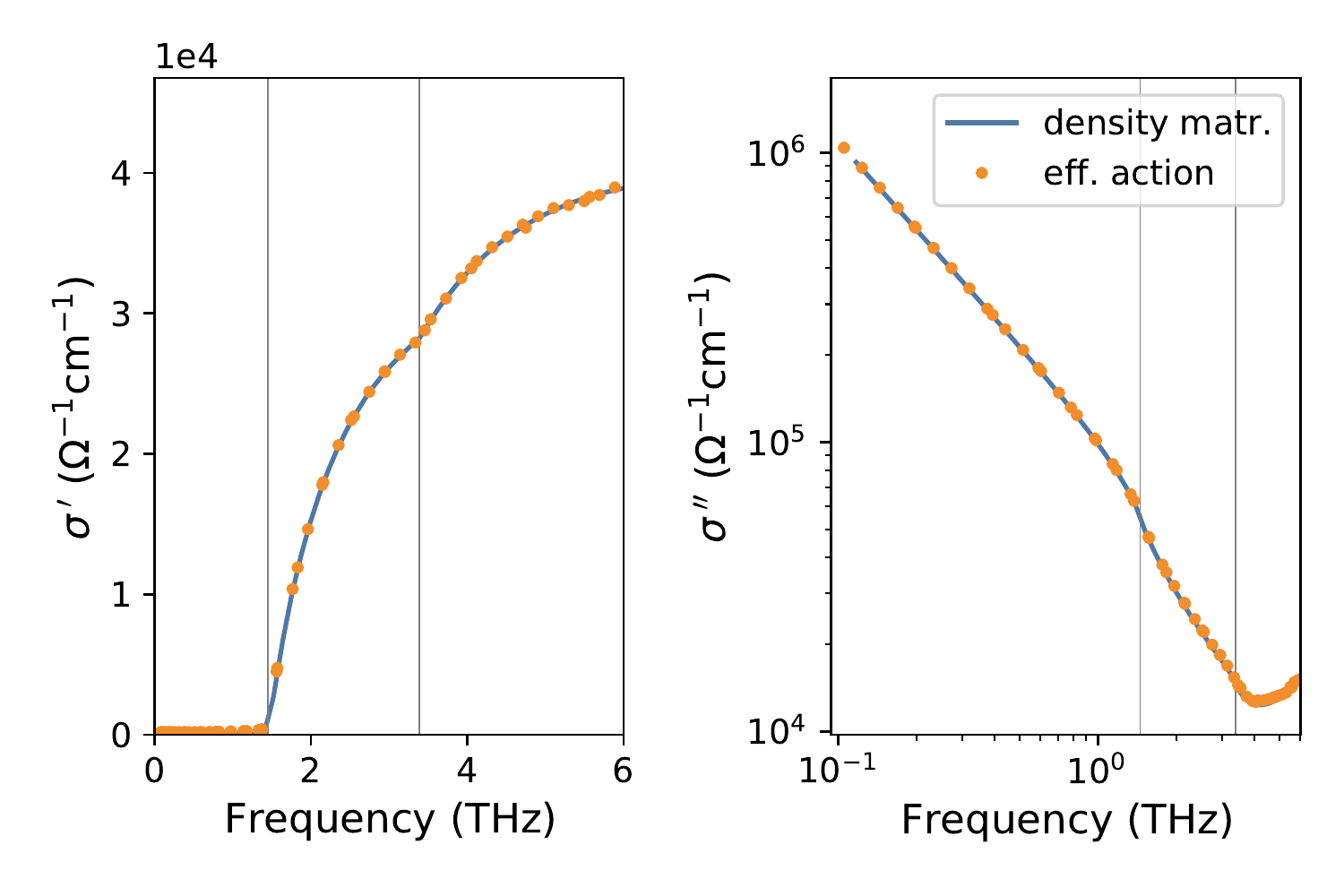}
    \caption{Real and imaginary part of optical conductivity computed in the time-dependent density matrix formalism (blue lines) and from diagrams Fig.~\ref{fig:diagrams-j}(a,b) in the effective action approach.
    There is perfect agreement between the two methods.}
    \label{fig:sigma-lin-comp-action}
\end{figure}
The paramagnetic first order current $j_P\big|_1$ is represented by the diagram in Fig.~\ref{fig:diagrams-j}(a) and explicitly 
given by a functional derivative of Eq.~\eqref{eq:j2-qp-para}. After analytical continuation and MB substitution one arrives at
\begin{eqnarray}
j_P(-\omega)\big|_1 = A(\omega)\sum_i \frac{v_{F_i}^2}{3N_i(0)}  \int d\epsilon d\epsilon' W_i(\epsilon,\epsilon') \chi_i^{\sigma_0\sigma_0}(\omega,\epsilon,\epsilon') \,.
\end{eqnarray}
The diamagnetic  first order current $j_P\big|_1$ reads
\begin{equation}
    j_D(-\omega)\big|_1 = -A(-\omega) \sum_i \frac{e^2}{m_i} s_i 
    \sum_{\mathbf{k}\sigma} 
    \langle
    c_{i\mathbf{k}\sigma}^\dagger c_{i\mathbf{k}\sigma}
    \rangle\,,
\end{equation}
where we have used that
$
    \frac{\delta}{\delta A(\omega)} A^2(0)
    =
    \frac{\delta}{\delta A(\omega)} \int d\omega' A(-\omega')A(\omega') = 2 A(-\omega)
$. Note that the $\mathbf{k}$-sum does not vanish away from the Fermi surface and therefore strongly depends on
the numerical cutoff. Here, we follow Murotani \cite{Murotani2019} and regularize the integral as
\begin{align}
    j_D(-\omega)\big|_1 = A(\omega) 
    \sum_i \frac{e^2 n_i}{m_i}  
    \int d\epsilon d\epsilon' 
    \frac{f(\epsilon)-f(\epsilon')}{\epsilon-\epsilon'}
    W_i(\epsilon,\epsilon')
\end{align}
with and Fermi function $f(\epsilon)$ and the band specific carrier density $n_i=k_{F_i}^3/3\pi^2$.

\section{Third Harmonic Generation}
\label{apdx:thg}
\begin{figure}[H]
    \centering
    \includegraphics[width=0.75\columnwidth]{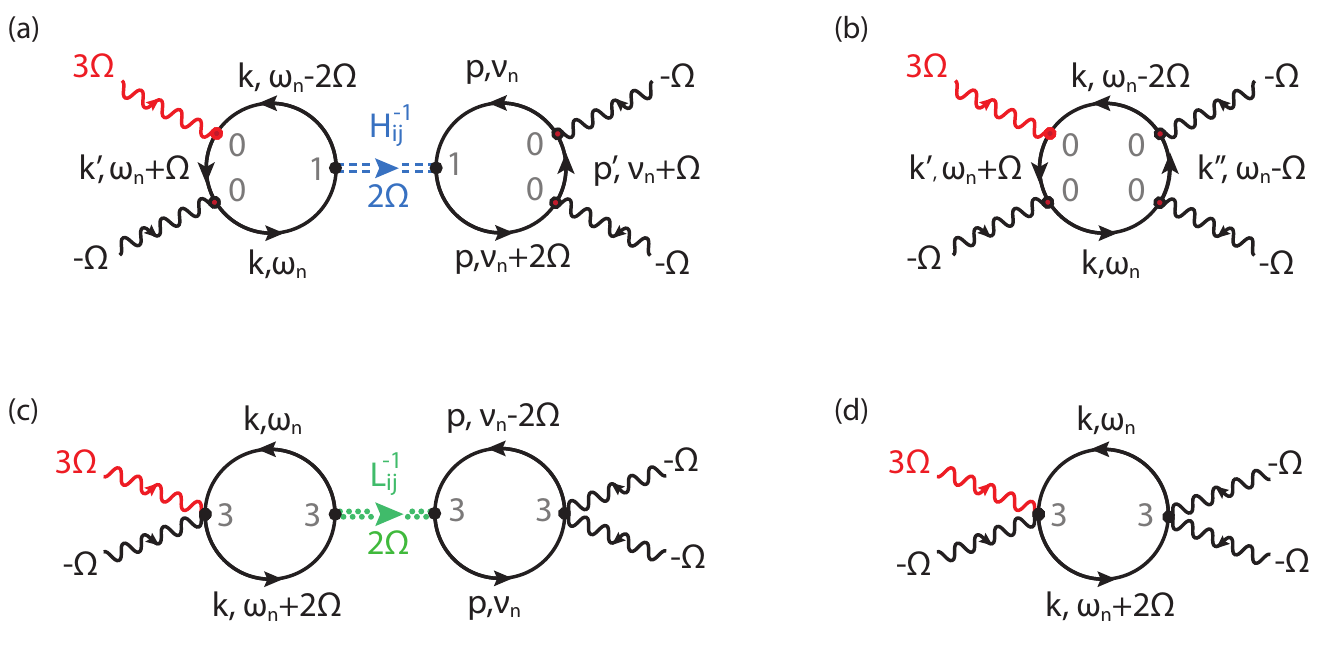}
    \caption{Diagrammatic representation of THG signal. Red photon legs denote $A$ with respect to which
    the functional derivative has been performed. Higgs and Leggett propagators (double lines) correspond to an RPA summation shown in Fig.~\ref{fig:diagrams-RPA}.}
    \label{fig:THG-diagram}
\end{figure}

For THG experiments, we are interested in the non-linear current $j(3\Omega)$ evaluated at $\omega=3\Omega$ where $\Omega$ is the dominant frequency
of the optical pulse $A(\Omega)$:
\begin{equation}
    j(-3\Omega) = - \frac{\delta S[A]}{\delta A(\omega)} \bigg|_{\omega=3\Omega} \,.
    \label{eq:THG-f}
\end{equation}
A diagrammatic representation of Eq.~\eqref{eq:THG-f} is shown in Fig.~\ref{fig:THG-diagram}. The field $A(\omega)$
with respect to which the functional derivative is performed is colored red. All four choices are equivalent.
The functional derivative forces the external frequency of the field $A$ to be $3\Omega$. In principal one
now needs to integrate over all remaining external frequencies, while satisfying energy conservation.
This can be numerically challenging. Here, we focus instead on the case of a monochromatic field
$A(t)=A_0 \cos \Omega t, \, A(\omega) = \frac{A_0}{2}\left(\delta(\omega-\Omega)+\delta(\omega+\Omega)\right)$ where external fields possess two discrete frequencies $\pm\Omega$.
Then energy conservation dictates all remaining external legs
 to carry frequency $-\Omega$. Note that the energy flow through collective Higgs or Leggett propagators is $2\Omega$,
 i.e. THG probes the optical kernel at twice the driving frequency as expected for a non-linear process.
 
\begin{figure}
    \centering
    \includegraphics[width=\columnwidth]{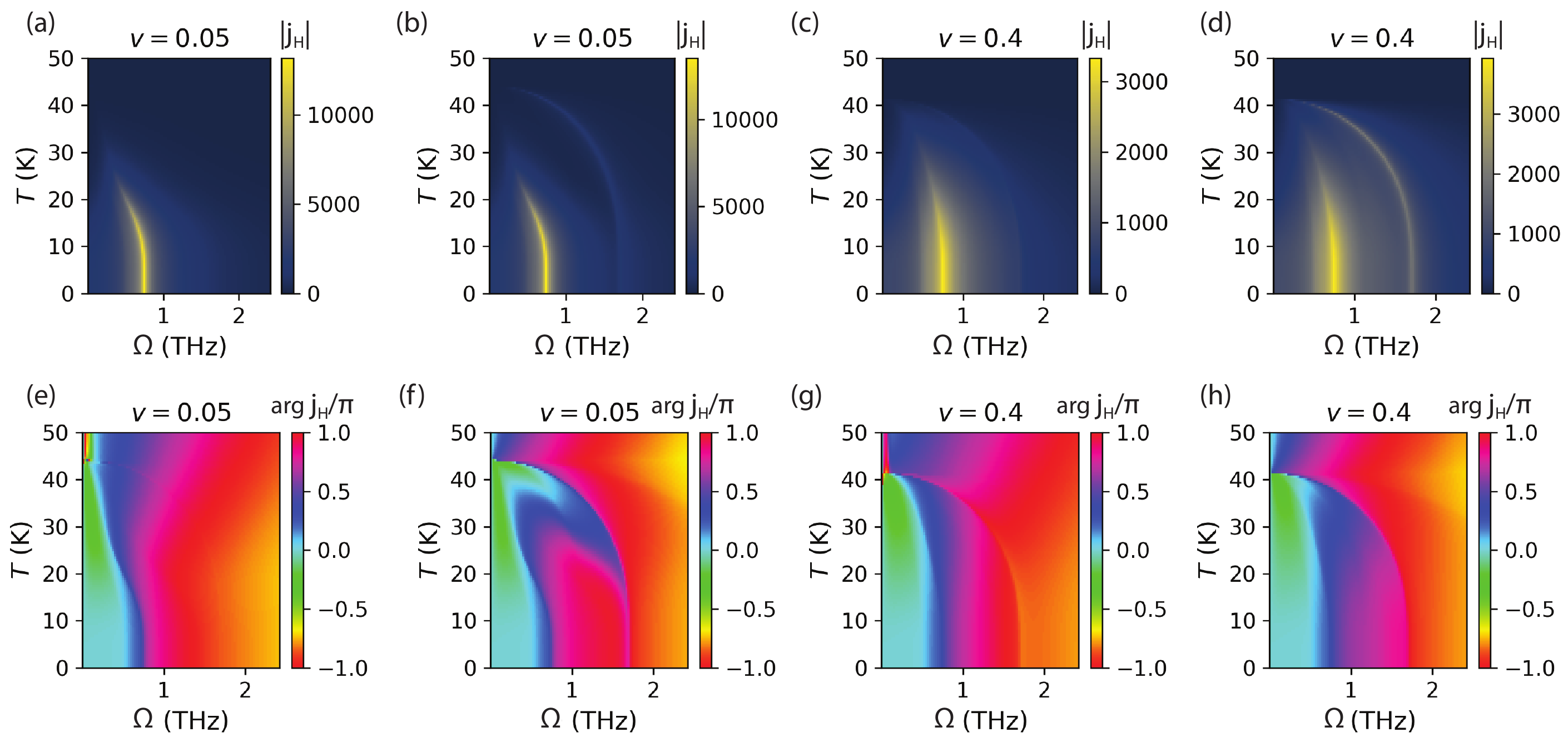}
    \caption{Magnitude, up to a prefactor, (a-d) and phase (e-h) of Higgs contribution to THG current as a function of driving frequency $\Omega$ and temperature $T$.
    (a),(b),(e),(f) correspond to the dirty-clean case with $\gamma_\pi=\SI{100}{\milli\electronvolt},\gamma_\sigma=\SI{0.01}{\milli\electronvolt}$ and 
    (c,d,g,h) correspond to the dirty-dirty case with $\gamma_\pi=\SI{100}{\milli\electronvolt},\gamma_\sigma=\SI{50}{\milli\electronvolt}$.    }
    \label{fig:j3-higgs}
\end{figure} 
\begin{figure}
    \centering
    \includegraphics[width=0.75\columnwidth]{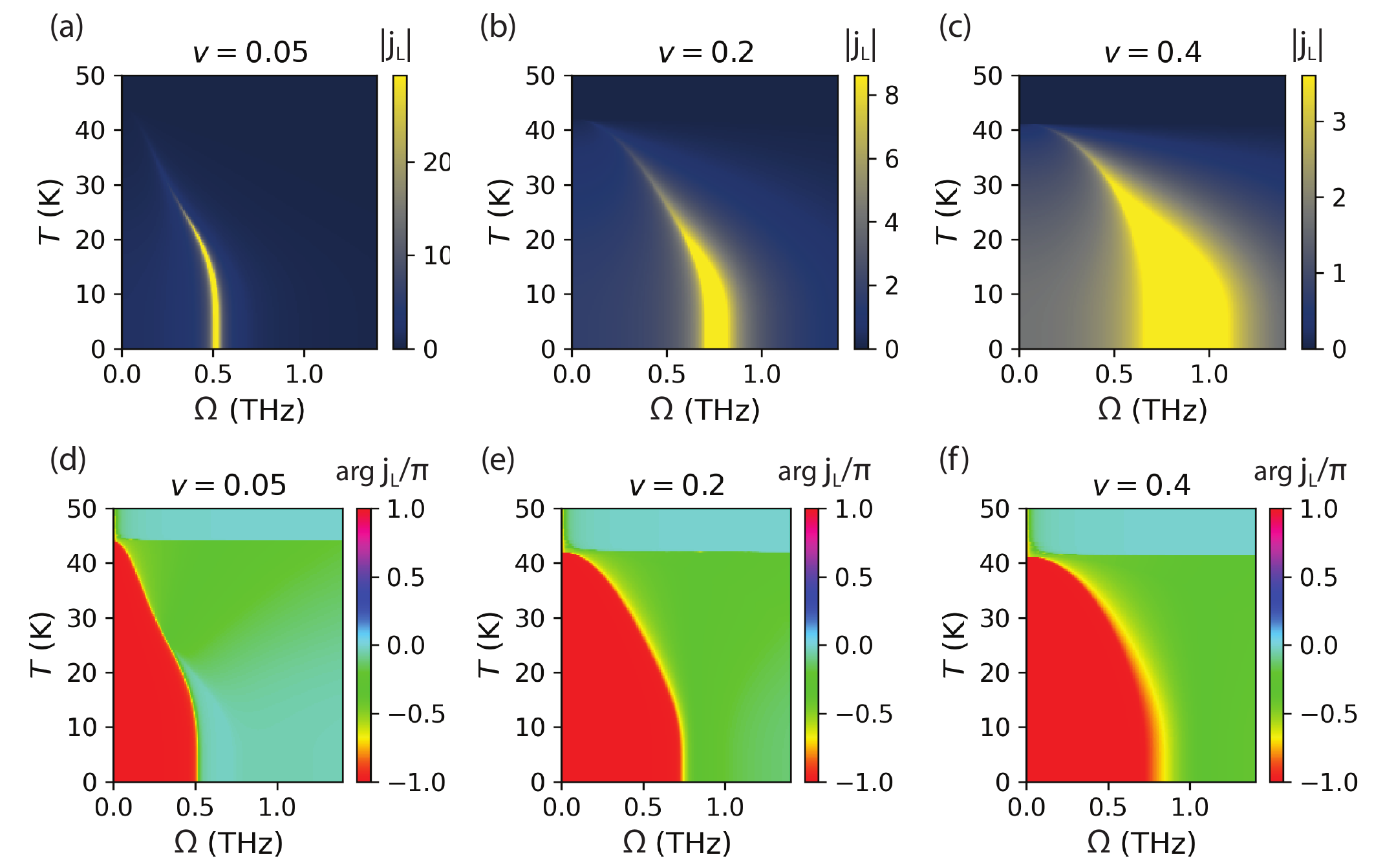}
    \caption{Magnitude, up to a prefactor, (a-c) and phase (d-f) of Leggett contribution to THG current as a function of driving frequency $\Omega$ and temperature $T$
    for various interband coupling parameters $v$ as denoted in plot titles.}
    \label{fig:j3-leggett}
\end{figure} 
\begin{figure}
    \centering
    \includegraphics[width=\columnwidth]{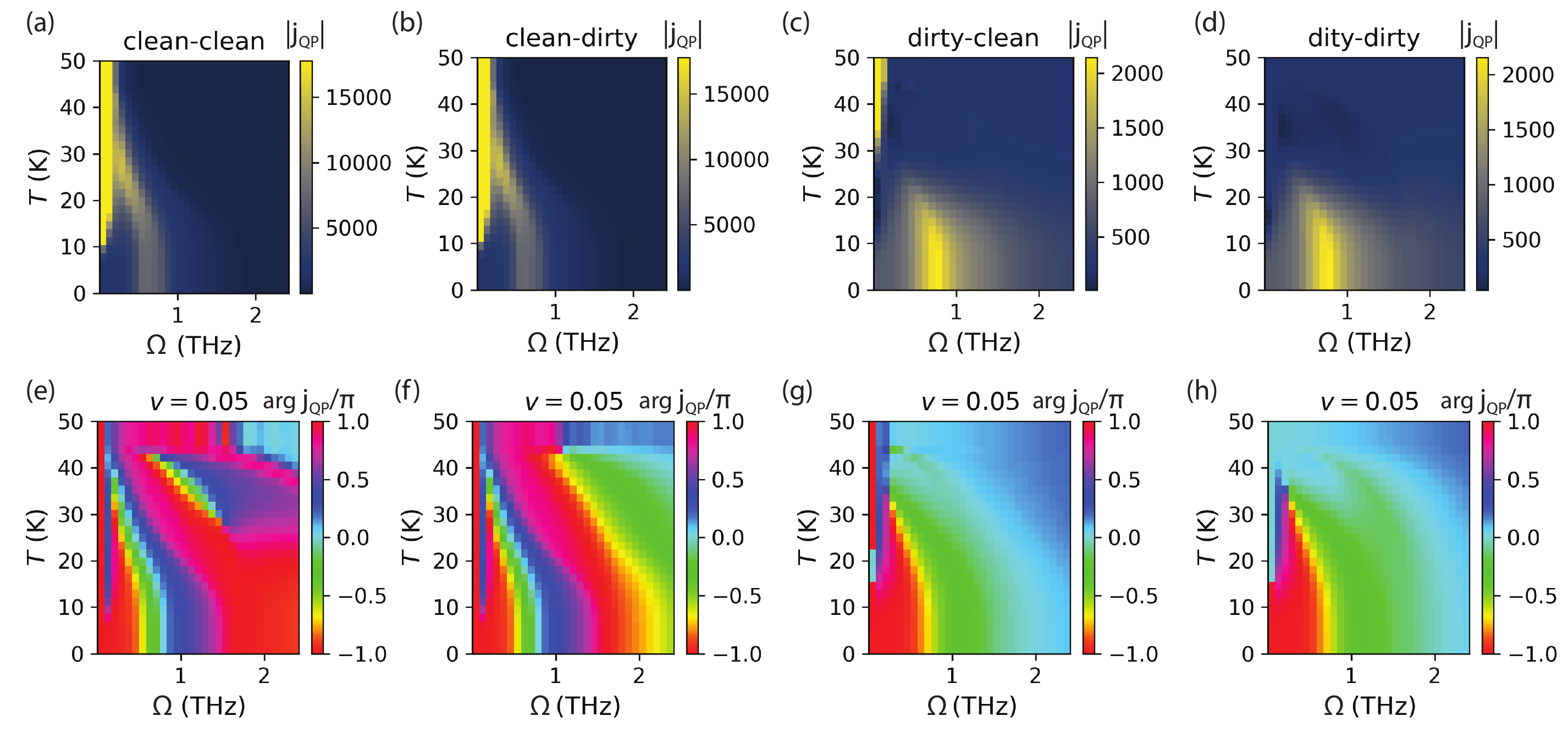}
    \caption{Magnitude, up to a prefactor, (a-d) and phase (e-h) of quasiparticle contribution to THG current as a function of driving frequency $\Omega$ and temperature $T$
    for cases (a) clean-clean (b) clean-dirty case, (c) dirty-clean (d) dirty-dirty.}
    \label{fig:j3-qp}
\end{figure} 

Fig.~\ref{fig:j3-higgs} shows magnitude and phase of the Higgs contribution to the THG current $j_H(3\Omega)$
as a function of $\Omega$ and $T$ for two interband couplings $v=0.05,0.4$. Panels (a,b,e,f) correspond to the limit of 
a dirty $\pi$-band and a clean $\sigma$-band, whereas the remaining panels are computed for two dirty bands. Both
cases are possible descriptions of MgB$_2$. Yellow spectral lines map out the Higgs resonance that follow 
$2\Delta_\pi, 2\Delta_\sigma$. In all cases the $\pi$-resonance is dominant, although the relative 
$\sigma$-contribution is
enhanced in the dirty-dirty limit and for strong $v$. 
Increased interband coupling $v$ decreases and broadens the overall Higgs response.

The Higgs resonance is sharp at small $v$, but much broader in the $v=0.4$ case. 
Therefore, slices along the $T$-axis for a given drive frequency $\Omega$ do not
exhibit a pronounced resonance peak. The observation of a resonance peak in Ref.~\cite{Kovalev2020} 
when experimentally sweeping the temperature would be therefore suggestive of a small $v$ coupling in MgB$_2$.
This is in disagreement to Refs. \cite{Blumberg2007, Giorgianni2019} that experimentally determined a large $v$ based on
evidence of the Leggett mode above $2\Delta_\pi$.

Lower panels in Fig.~\ref{fig:j3-higgs} show a phase jump of $\pi$ in the THG current across the first Higgs resonance along
the $\Omega$ direction that is most pronounced at low temperatures. The phase also shows features of the 
$\sigma$-Higgs resonance, albeit less clearly. Approaching the resonance along the $T$ axis does not yield
a phase behavior that is consistently simple to interpret.
These results are to be contrasted to the clean case where one expects a phase jump of $\pi/2$.

Fig.~\ref{fig:j3-leggett} shows the amplitude and phase response of the Leggett THG signal for three 
different coupling strengths $v=0.02,0.2,0.5$. The overall contribution is about three magnitudes smaller
than the Higgs contribution and therefore negligible. At large coupling, the Leggett resonance is 
very broad but sharpens at high temperatures. This observation was first reported in Ref.~\cite{Murotani2017}.
The phase shows a clear $\pi/2$-jump across the resonance for all temperatures below $T_C$.

The quasiparticle contribution is shown in Fig.~\ref{fig:j3-qp} for $v=0.05$ in different impurity cases. Here \textit{clean} refers
to $\gamma=\SI{0.01}{\milli\electronvolt}$ and \textit{dirty} specifies $\gamma=\SI{100}{\milli\electronvolt}$. Results at different $v$ are nearly identical
since the only $v$-dependent quantity in Eq.~\eqref{eqn:s4qp-para} is the superconducting order parameter at finite $T$. For all
impurity concentrations, the quasiparticle THG signal is peaked at the onset of the quasiparticle continuum of the $\pi$-band. 
The signal is about one order of magnitude smaller than the Higgs contribution in the small $v$ case. For $v=0.4$, the quasiparticle signal
remains nearly identical but the Higgs signal increases, so that the Higgs contribution is only slightly larger.
In all but the \textit{dirty-dirty} case, the quasiparticle signal has a large contribution for small $\Omega$ and large $T$.

The bottom row of Fig.~\ref{fig:j3-qp} shows the phase of the non-linear THG signal. In the \textit{dirty-clean} and \textit{dirty-dirty} cases we observe a clear phase jump of $\pi/2$ across a resonance.

\section{Density matrix equation of motion formulation}
\label{apdx:model}
The following derivation uses notation similar to \cite{Murotani2019} and is included to make the present work self-contained. We start from the mean field BCS Hamiltonian
\begin{eqnarray}
  \mathcal{H}_0 = \sum_{i\mathbf{k}\sigma}  
	\epsilon_{i\mathbf{k}}c_{i\mathbf{k}\sigma}^\dagger
	c_{i\mathbf{k}\sigma} + \sum_{i\mathbf{k}}^{}
	\left( \Delta_i c_{i\mathbf{-k}\uparrow }^\dagger
	c_{i\mathbf{k}\downarrow }^\dagger  \right) \,,
\end{eqnarray}
where $\epsilon_{i\mathbf{k}} = s_i \left(\mathbf{k}^2/2m_i -
\epsilon_{F_i}\right)$ and the superconducting order parameter is self-consistently determined by 
$\Delta_i = \sum_{j\mathbf{k}}^{}U_{ij} \langle c_{j-\mathbf{k}\downarrow
}c_{j\mathbf{k}\uparrow }\rangle$.
To incorporate the pulse, we add paramagnetic and diamagnetic coupling to the laser field through
\begin{eqnarray}
	\mathcal{H}_1 = -\sum_{i\mathbf{kk'}\sigma}^{}
	\mathbf{J}_{i\mathbf{kk'}} \cdot \mathbf{A} \,
	c_{i\mathbf{k}\sigma}^\dagger  c_{i\mathbf{k}'\sigma} +
	\sum_{i\mathbf{k}\sigma}^{} \frac{s_i e^2}{2m_i} \mathbf{A}^2 \,
	c_{i\mathbf{k}\sigma}^\dagger c_{i\mathbf{k}\sigma}.
\end{eqnarray}
The full Hamiltonian is given by $\mathcal{H} = \mathcal{H}_0 +\mathcal{H}_1$. The current density can be calculated using 
\begin{eqnarray}
  \mathbf{j} = - \bigg\langle \frac{\delta\mathcal{H}}{\delta
  \mathbf{A}}\bigg\rangle = \mathbf{j}_P + \mathbf{j}_D,
\end{eqnarray}
where the paramagnetic component $\mathbf{j}_P$ and diamagnetic component $\mathbf{j}_D$ correspond to the first and second terms in $\mathcal{H}_1$ respectively. The equilibrium Hamiltonian can be diagonalized through a Bogoliubov transformation, where we introduce the fermionic Bogoliubov quasiparticles in the form of the two-component spinor
\begin{eqnarray}
  \VV{\psi_{i\mathbf{k}}^1}{\psi_{i\mathbf{k}}^2} = \MM{u_{i \mathbf{k} }}{-v_{i \mathbf{k}}}{v^*_{i \mathbf{k}}}{u_{i \mathbf{k}}} \VV{c_{i \mathbf{k} \uparrow}}{c^\dagger_{i (-\mathbf{k}) \downarrow}},
\end{eqnarray}
where
\begin{eqnarray}
  E_{i\vk} = \sqrt{\epsilon^2_{i\vk}+|\Delta_i^{\text{eq}}|^2}, \quad
  u_{i\vk}^2 = \frac{1}{2}\left( 1 + \frac{\epsilon_{i\vk}}{E_{i\vk}} \right), \quad
  |v_{i\vk}|^2 = \frac{1}{2}\left( 1 - \frac{\epsilon_{i\vk}}{E_{i\vk}} \right) ,
\end{eqnarray}
and $\Delta_i^{\text{eq}}$ is the equilibrium value of the gap. We are free to choose the initial phase so that $u_{i\mathbf{k}}$ and $v_{i\mathbf{k}}$ are both real. Next, we construct the density matrix $\rho$ in the Bogoliubov quasiparticle basis
\begin{eqnarray}
  \rho = \ket{\psi_0}\bra{\psi_0} = 
  \begin{pmatrix}
    \rho_{i\mathbf{kk'}}^{11} &
    \rho_{i\mathbf{kk'}}^{12} \\
    \rho_{i\mathbf{kk'}}^{21} &
    \rho_{i\mathbf{kk'}}^{22}
  \end{pmatrix}
  =
  \begin{pmatrix}
    \langle \psi^{1\dagger}_{i\mathbf{k}} \psi^{1}_{i\mathbf{k'}}\rangle
    &
    \langle \psi^{1\dagger}_{i\mathbf{k}} \psi^{2}_{i\mathbf{k'}} \rangle
    \\
    \langle 
    \psi^{2\dagger}_{i\mathbf{k}} \psi^{1}_{i\mathbf{k'}}
    \rangle
    &
    \langle 
    \psi^{2\dagger}_{i\mathbf{k}} \psi^{2}_{i\mathbf{k'}} 
    \rangle
  \end{pmatrix}.
\end{eqnarray}
We can now rewrite the Hamiltonian in terms of the quasiparticles and their expectation values
\begin{align}
    \mathcal{H}_0 &=  \sum_{i\mathbf{k}}\VVT{\psi_{i\mathbf{k}}^{1\dagger}}{\psi_{i\mathbf{k}}^{2\dagger}}
    \MM{
    \epsilon_{i \textbf{k}}(u_{i\textbf{k}}^2 - v_{i\textbf{k}}^2) + u_{i\textbf{k}}v_{i\textbf{k}}(\Delta_i + \Delta_i^*)}
    {2u_{i\textbf{k}}v_{i\textbf{k}}\epsilon_{i \textbf{k}} + \Delta_i^* v_{i\textbf{k}}^2 - \Delta_i u_{i\textbf{k}}^2}
    {2u_{i\textbf{k}}v_{i\textbf{k}}\epsilon_{i \textbf{k}} + \Delta_i v_{i\textbf{k}}^2 - \Delta_i^* u_{i\textbf{k}}^2}
    {-\epsilon_{i \textbf{k}}(u_{i\textbf{k}}^2 - v_{i\textbf{k}}^2) - u_{i\textbf{k}}v_{i\textbf{k}}(\Delta_i + \Delta_i^*)}
    \VV{\psi_{i\mathbf{k}}^1}{\psi_{i\mathbf{k}}^2} .
\end{align}
In equilibrium, where $\Delta_i = \Delta_i^{\text{eq}}$, this reduces to 
\begin{align}
    \mathcal{H}_0^{\text{eq}} &= \sum_{i\mathbf{k}}\VVT{\psi_{i\mathbf{k}}^{1\dagger}}{\psi_{i\mathbf{k}}^{2\dagger}}
    \MM{E_{i\mathbf{k}}}{0}{0}{-E_{i\mathbf{k}}}
    \VV{\psi_{i\mathbf{k}}^1}{\psi_{i\mathbf{k}}^2}.
\end{align}
In non-equilibrium, we can write the time-dependent gap as $\Delta_i(t) = \Delta_i^{\text{eq}} + \delta \Delta_i (t)$. Furthermore we can decompose $\delta \Delta_i (t) = \delta \Delta_i' (t) + i \delta \Delta_i'' (t)$ into real and imaginary parts. Then, we find
\begin{align}
    \mathcal{H}_0 &= \mathcal{H}_0^{\text{eq}} + 
    \sum_{i\mathbf{k}}\VVT{\psi_{i\mathbf{k}}^{1\dagger}}{\psi_{i\mathbf{k}}^{2\dagger}}
    \left[ \delta \Delta_i'
	\begin{pmatrix}
	2u_{i\mathbf{k}}v_{i\mathbf{k}}& -u_{i\mathbf{k}}^2 + v_{i\mathbf{k}}^2 \\
	-u_{i\mathbf{k}}^2 + v_{i\mathbf{k}}^2 & -2u_{i\mathbf{k}}v_{i\mathbf{k}}
	\end{pmatrix}
	+ \delta \Delta_i''
	\begin{pmatrix}
	0&-i\\
	i&0
	\end{pmatrix}\right]
    \VV{\psi_{i\mathbf{k}}^1}{\psi_{i\mathbf{k}}^2}.
\end{align}
The coupling to the pulse becomes
\begin{align}
    \mathcal{H}_1 = &- \sum_{i\mathbf{k}\mathbf{k}'} \textbf{J}_{i\mathbf{k}\mathbf{k}'}\cdot \textbf{A}
	\VVT{\psi_{i\mathbf{k}}^{1\dagger}}{\psi_{i\mathbf{k}}^{2\dagger}}
	\begin{pmatrix}
	l_{i\mathbf{k}\mathbf{k}'} & -p_{i\mathbf{k}\mathbf{k}'}\\
	p_{i\mathbf{k}\mathbf{k}'} & l_{i\mathbf{k}\mathbf{k}'}
	\end{pmatrix}
	\VV{\psi_{i\mathbf{k}'}^1}{\psi_{i\mathbf{k}'}^2} \nonumber \\
	&+ 
	\sum_{i\mathbf{k}} s_i \frac{e^2 \textbf{A}^2}{2m_iE_{i\mathbf{k}}}
    \VVT{\psi_{i\mathbf{k}}^{1\dagger}}{\psi_{i\mathbf{k}}^{2\dagger}}
	\begin{pmatrix}
	\epsilon_{i\mathbf{k}} & \Delta_i^{\text{eq}}\\
	\Delta_i^{\text{eq}} & -\epsilon_{i \mathbf{k}}
	\end{pmatrix}
    \VV{\psi_{i\mathbf{k}}^1}{\psi_{i\mathbf{k}}^2},
\end{align}
where
\begin{align}
	\nonumber l_{i\mathbf{k}\mathbf{k}'} &=u_{i\mathbf{k}}u_{i\mathbf{k}'} + v_{i\mathbf{k}}v_{i\mathbf{k}'}\,,\\
	p_{i\mathbf{k}\mathbf{k}'} &= v_{i\mathbf{k}}u_{i\mathbf{k}'} - u_{i\mathbf{k}}v_{i\mathbf{k}'} .
	\end{align}
Notice the total Hamiltonian can be expressed as
\begin{align}
    \mathcal{H}_0 + \mathcal{H}_1 &= \sum_{ab}\sum_{i\mathbf{kk}'} \psi_{i \mathbf{k}}^{a\dagger} h_{i\mathbf{kk}'}^{ab}\psi_{i\mathbf{k}'}^b .
\end{align}
Next, we decompose the gap into real and imaginary parts $\Delta_i = \Delta_i' + i\Delta_i''$. After the transformation, the gap equation becomes
\begin{align}
    \Delta_{i}' &= \sum_j U_{ij} \sum_{\textbf{k}} 
    \left[ -u_{j\textbf{k}} v_{j\textbf{k}} \left( \rho_{j\mathbf{kk}}^{11} - \rho_{j\mathbf{kk}}^{22} \right) 
    + 
    \frac{1}{2} \left( u_{j\textbf{k}}^2 - v_{j\textbf{k}}^2\right) \left(\rho_{j\mathbf{kk}}^{21} + \rho_{j \mathbf{kk}}^{12} \right) \right]\,,\\
	\Delta_i'' &= \sum_j U_{ij} \sum_{\textbf{k}} \frac{1}{2i} \left(\rho_{j \mathbf{kk}}^{21} - \rho_{j\mathbf{kk}}^{12} \right) .
\end{align}
Finally, we write the paramagnetic and diamagnetic current densities in terms of the quasiparticles
\begin{align}
    \mathbf{j}_P &= \sum_{i\mathbf{k}\mathbf{k}'}\mathbf{J}_{i\mathbf{k}\mathbf{k}'}
    \left[l_{i\mathbf{k}\mathbf{k}'} \left( \rho_{i\mathbf{kk}'}^{11} + \rho_{i\mathbf{kk}'}^{22} \right) + p_{i\textbf{k}\textbf{k}'} \left( \rho_{i\mathbf{kk}'}^{21} - \rho_{i\mathbf{kk}'}^{12} \right)\right] 
    \label{eqn:jp}\,,\\
    \mathbf{j}_D &= 
    - \sum_{i\mathbf{k}} s_i \frac{e^2 \mathbf{A}}{m_i E_{i\mathbf{k}}} \left[ \left( u_{i\textbf{k}}^2 - v_{i\textbf{k}}^2\right) \left( \rho_{i\mathbf{kk}}^{11} - \rho_{i\mathbf{kk}}^{22} \right) +2u_{i\textbf{k}}v_{i\textbf{k}} \left(\rho_{i\mathbf{kk}}^{21} + \rho_{i\mathbf{kk}}^{12} \right) \right] \label{eqn:jd} .
\end{align}
We use Heisenberg's equation of motion to find the time dependence of the quasiparticle expectation values. Writing the density matrix in vector form,
\begin{eqnarray}
  \rho_{i\mathbf{kk}'} = 
  \begin{pmatrix}
    \rho_{i\mathbf{kk'}}^{11} \\
    \rho_{i\mathbf{kk'}}^{12} \\
    \rho_{i\mathbf{kk'}}^{21} \\
    \rho_{i\mathbf{kk'}}^{22}
  \end{pmatrix} \,,
\end{eqnarray}
Heisenberg's equation of motion is stated as
\begin{eqnarray}
  i\partial_t \rho_{i\mathbf{k}\mathbf{k}'}
  =
  \sum_{\mathbf{q}}^{}
  \left[ 
    H_{i\mathbf{k}' \mathbf{q}}^{(1)} \rho_{i\mathbf{kq}} -
    H_{i\mathbf{qk}}^{(2)}
    \rho_{i\mathbf{qk'}}
  \right] \label{eqn:eom} ,
\end{eqnarray}
where we have defined the following two matrices:
\begin{eqnarray}
  H_{i\mathbf{kk'}}^{(1)}
  =
  \begin{pmatrix}
    h_{i\mathbf{kk'}}^{11} & h_{i\mathbf{kk'}}^{12} & & \\
    h_{i\mathbf{kk'}}^{21} & h_{i\mathbf{kk'}}^{22} & & \\
    && h_{i\mathbf{kk'}}^{11} & h_{i\mathbf{kk'}}^{12} \\
    && h_{i\mathbf{kk'}}^{21} & h_{i\mathbf{kk'}}^{22} & 
  \end{pmatrix}
,\quad
  H_{i\mathbf{kk'}}^{(2)}
  =
  \begin{pmatrix}
    h_{i\mathbf{kk'}}^{11} && h_{i\mathbf{kk'}}^{21} & \\
    & h_{i\mathbf{kk'}}^{11} && h_{i\mathbf{kk'}}^{21} \\
    h_{i\mathbf{kk'}}^{12} && h_{i\mathbf{kk'}}^{22} & \\
    & h_{i\mathbf{kk'}}^{12} && h_{i\mathbf{kk'}}^{22}
  \end{pmatrix}\,.
\end{eqnarray}
Note that we work in natural units where $\hbar=1$.
In order to solve the equation of motion with the impurity scattering replacement, we proceed using a perturbative approach with respect to $\mathbf{A}$. This involves expanding all relevant quantities in powers of $\mathbf{A}$. For instance, we expand $\rho_{i\mathbf{kk}'}$ as $\rho_{i\mathbf{kk}'} = \rho_{i\mathbf{kk}'} \big|_0 + \rho_{i\mathbf{kk}'} \big|_1 + \rho_{i\mathbf{kk}'} \big|_2 + \dots$, where $\rho_{i\mathbf{kk}'} \big|_1$ is proportional to $\mathbf{A}$, $\rho_{i\mathbf{kk}'} \big|_2$ is proportional to $\mathbf{A}^2$, etc.

We now solve each order separately. The zeroth-order components are simply the equilibrium values, where the quasiparticles are occupied according to Fermi statistics
\begin{eqnarray}
  \rho_{i\mathbf{kk'}}\big|_0 = \delta_{kk'} 
  \begin{pmatrix}
    f_{i\mathbf{k}}
    \\
    0
    \\
    0
    \\
    1-f_{i\mathbf{k}}
  \end{pmatrix}\,,
\end{eqnarray}
\begin{eqnarray}
  H_{i\mathbf{kk'}}^{(1)}\big|_0
  =
  \delta_{\mathbf{kk'}}E_{i\mathbf{k}}
  \begin{pmatrix}
    1 & & & \\
     & -1 & & \\
    && 1 &  \\
    &&  & -1 & 
  \end{pmatrix}
, \quad
  H_{i\mathbf{kk'}}^{(2)}\big|_0
  =
  \delta_{\mathbf{kk'}}E_{i\mathbf{k}}
  \begin{pmatrix}
    1 && & \\
    & 1 &&  \\
     && -1 & \\
    & && -1
  \end{pmatrix}\,.
\end{eqnarray}
We can also use these values to calculate the equilibrium value of the gap, exchanging the sum in the gap equation with an integral over the energy:
\begin{align}
\Delta_i^{\text{eq}} = \sum_{j}U_{ij}N_j(0)\Delta_j^{\text{eq}} \int_{-\omega_d}^{\omega_d} \frac{d\epsilon}{2\sqrt{\epsilon^2 + (\Delta_i^{\text{eq}})^2}} \tanh \left(\frac{\beta}{2} \sqrt{\epsilon^2 + (\Delta_i^\text{eq})^2}\right)\,.
\end{align}
Now we proceed with the first order. Considering only the terms proportional to $\mathbf{A}$, the equation of motion becomes
\begin{eqnarray}
i\partial_t \rho_{i\mathbf{k}\mathbf{k}'}\big|_1
  =
  \left( 
    H_{i\mathbf{k}' \mathbf{k'}}^{(1)} \big|_0
    -
    H_{i\mathbf{kk}}^{(2)}\big|_0
  \right)
    \rho_{i\mathbf{kk'}}\big|_1
  +
  \left(
    H_{i\mathbf{k}' \mathbf{k}}^{(1)}\big|_1 \rho_{i\mathbf{kk}}\big|_0 
    -
    H_{i\mathbf{k'k}}^{(2)}\big|_1
    \rho_{i\mathbf{k'k'}}\big|_0
  \right) ,
\end{eqnarray}
where
\begin{eqnarray}
  H_{i\mathbf{kk'}}^{(1)}\big|_1
  =
  -\mathbf{J}_{i\mathbf{kk'}} \cdot \mathbf{A} \,
  \begin{pmatrix}
    l_{i\mathbf{kk'}} & -p_{i\mathbf{kk'}} & & \\
    p_{i\mathbf{kk'}} & l_{i\mathbf{kk'}} & & \\
    && l_{i\mathbf{kk'}} & -p_{i\mathbf{kk'}} \\
    &&  p_{i\mathbf{kk'}}& l_{i\mathbf{kk'}}
  \end{pmatrix} ,
\end{eqnarray}
\begin{eqnarray}
  H_{i\mathbf{kk'}}^{(2)}\big|_1
  =
  -\mathbf{J}_{i\mathbf{kk'}} \cdot \mathbf{A} \,
  \begin{pmatrix}
    l_{i\mathbf{kk'}} && p_{i\mathbf{kk'}} & \\
    & l_{i\mathbf{kk'}} && p_{i\mathbf{kk'}} \\
    -p_{i\mathbf{kk'}} && l_{i\mathbf{kk'}} & \\
    & -p_{i\mathbf{kk'}}  &&l_{i\mathbf{kk'}} 
  \end{pmatrix}.
\end{eqnarray}
We solve the equation by writing it in terms of new functions $F_{i\mathbf{kk'}}^{ab} $. The first order expression of $\rho_{i\mathbf{kk}'}$ becomes
\begin{align}
    \rho_{i\mathbf{kk}'}\big|_1 = 
    \mathbf{J}_{i\mathbf{k'k}}\cdot\mathbf{e}
    \begin{pmatrix}
      l_{i\mathbf{kk'}} F_{i\mathbf{kk'}}^{11}\\
      p_{i\mathbf{kk'}} F_{i\mathbf{kk'}}^{12}\\
      -p_{i\mathbf{kk'}} F_{i\mathbf{kk'}}^{21}\\
      l_{i\mathbf{kk'}} F_{i\mathbf{kk'}}^{22}
    \end{pmatrix}
    \label{eqn:1storder} ,
\end{align}
where $F_{i}^{ab} $ is defined according to
\begin{align}
	\left[ i \hbar \frac{\partial}{\partial t} - (E'-E)\right]F_i^{11}(\epsilon, \epsilon') &= (f'-f)A \,,\\
	\left[ i \hbar \frac{\partial}{\partial t} - (E'+E)\right]F_i^{21}(\epsilon, \epsilon') &=- (1-f-f')A \,,\\
	F_i^{12}(\epsilon, \epsilon') &= F_i^{21}(\epsilon, \epsilon')^* \,,\\
	F_i^{22}(\epsilon, \epsilon') &= F_i^{11}(\epsilon, \epsilon')^* .
\end{align}
We have introduced a simplified notation $\epsilon = \epsilon_{i \textbf{k}}, E' = E_{i\textbf{k}'}, F_{i\mathbf{kk'}}^{ab}=F_i^{ab}(\epsilon,\epsilon')$ etc. The collective modes do not couple linearly to light, and correspondingly $\delta \Delta_i\big|_1 =0$. This can be confirmed by using Eq.~\eqref{eqn:1storder} in the gap equation. Hence, we only examine the response of the current. By substituting Eq.~\eqref{eqn:1storder} into Eq. (\ref{eqn:jp}), changing the momentum sums into integrals over energy, and making the Mattis-Bardeen replacement (\ref{eq:MB}), we find
\begin{equation}
	\mathbf{j}_P\big|_1 = \mathbf{e} \sum_{i}\frac{e^2 n_i}{m_i} \int d\epsilon d\epsilon' W_i(\epsilon, \epsilon') \left[ l_i(\epsilon, \epsilon')^2 \Re F_i^{11}(\epsilon, \epsilon') + p_i(\epsilon,\epsilon')^2 \Re F_i^{21}(\epsilon,\epsilon')\right]
	\label{eqn:jp1} .
	\end{equation}
We can also derive the induced diamagnetic current,
\begin{equation}
	\textbf{j}_\text{D}(t) \bigg|_1 = -\textbf{A}\sum_{i} \frac{e^2 n_i}{m_i}
	\label{eqn:jd1} .
\end{equation}
This derivation is carefully discussed in \cite{Murotani2019}.

We now consider the second order solution. Keeping only terms proportional to $\mathbf{A}$, we find that the off-diagonal terms in the equation of motion vanish, and as a result (\ref{eqn:eom}) becomes
\begin{eqnarray}
  i \partial_t \rho_{i\mathbf{k}\mathbf{k}}
  &=&
    \left( 
      H_{i\mathbf{kk}}^{(1)}  - H_{i\mathbf{kk}}^{(2)} 
    \right)\bigg|_0
    \rho_{i\mathbf{kk}}\big|_2
    +
    \sum_{\mathbf{q}}
    \left( 
    H_{i\mathbf{kq}}^{(1)}\big|_1 \rho_{i\mathbf{kq}}\big|_1 -
    H_{i\mathbf{qk}}^{(2)}\big|_1 \rho_{i\mathbf{qk}}\big|_1
    \right)
    +
    \left( 
      H_{i\mathbf{kk}}^{(1)} - H_{i\mathbf{kk}}^{(2)} 
    \right)\bigg|_2
      \rho_{i\mathbf{kk}}\big|_0 .
\end{eqnarray}
We can decompose $H^{1}_{i\mathbf{kk}}\big|_2$ and $H^{2}_{i\mathbf{kk}}\big|_2$ into contributions from the diamagnetic quasiparticle current, Higgs mode, and Leggett mode as follows.
\begin{eqnarray}
  H_{i\mathbf{kk}}^{(1,2)}\big|_2 = 
  H_{i\mathbf{kk}}^{(1,2)}\big|_{2,D}
  +H_{i\mathbf{kk}}^{(1,2)}\big|_{2,H}
  +H_{i\mathbf{kk}}^{(1,2)}\big|_{2,L} ,
\end{eqnarray}
where the diamagnetic quasiparticle current contribution is
\begin{eqnarray}
  H_{i\mathbf{kk'}}^{(1)}\big|_{2,D}
  =
  \delta_{\mathbf{kk'}} \frac{s_ie^2}{2m_iE_{i\mathbf{k}}}\mathbf{A}^2
  \begin{pmatrix}
    \epsilon_{i\mathbf{k}} & \Delta_i  & & \\
    \Delta_i & -\epsilon_{i\mathbf{k}} & & \\
    && \epsilon_{i\mathbf{k}} & \Delta_i \\
    && \Delta_i & -\epsilon_{i\mathbf{k}}
  \end{pmatrix},
\quad
  H_{i\mathbf{kk'}}^{(2)}\big|_{2,D}
  =
  \delta_{\mathbf{kk'}} \frac{s_ie^2}{2m_i}\mathbf{A}^2
  \begin{pmatrix}
    \epsilon_{i\mathbf{k}} && \Delta_i & \\
    & \epsilon_{i\mathbf{k}} && \Delta_i \\
     \Delta_i&& -\epsilon_{i\mathbf{k}} & \\
    & \Delta_i &&-\epsilon_{i\mathbf{k}} 
  \end{pmatrix},
\end{eqnarray}
the Higgs contribution is
\begin{eqnarray}
  H_{i\mathbf{kk'}}^{(1)}\big|_{2,H}
  =
  \delta_{\mathbf{kk'}} \frac{\delta \Delta_i'\big|_2}{E_{i\mathbf{k}}}
  \begin{pmatrix}
    \Delta_i & -\epsilon_{i\mathbf{k}} & & \\
    -\epsilon_{i\mathbf{k}} & -\Delta_i & & \\
    && \Delta_i & -\epsilon_{i\mathbf{k}} \\
    && -\epsilon_{i\mathbf{k}} &  -\Delta_i
  \end{pmatrix},
\quad
  H_{i\mathbf{kk'}}^{(2)}\big|_{2,H}
  =
  \delta_{\mathbf{kk'}} \delta \Delta_i'\big|_2
  \begin{pmatrix}
    \Delta_i && -\epsilon_{i\mathbf{k}} & \\
    & \Delta_i && -\epsilon_{i\mathbf{k}} \\
    -\epsilon_{i\mathbf{k}} && -\Delta_i & \\
    & -\epsilon_{i\mathbf{k}} && -\Delta_i
  \end{pmatrix},
\end{eqnarray}
and the Leggett contribution is
\begin{eqnarray}
  H_{i\mathbf{kk'}}^{(1)}\big|_{2,L}
  =
  \delta_{\mathbf{kk'}} \delta \Delta_i''\big|_2
  \begin{pmatrix}
     & -i & & \\
    i &  & & \\
    &&  & -i \\
    && i &  
  \end{pmatrix},
\quad
  H_{i\mathbf{kk'}}^{(2)}\big|_{2,L}
  =
  \delta_{\mathbf{kk'}} \delta \Delta_i''\big|_2
  \begin{pmatrix}
     && i & \\
    &  && i \\
    -i &&  & \\
    & -i && 
  \end{pmatrix}.
\end{eqnarray}
To simplify the equation of motion, we introduce a new angle-averaged quantity, $r_i^{ab}(\epsilon)$, defined by
\begin{align}
    r_i^{ab}(\epsilon) &= \int \frac{d \Omega_\textbf{k}}{4 \pi} \rho_{i\mathbf{kk}}^{ab}\big|_2 .
\end{align}
The equations of motion can then be written as follows
\begin{equation}
	i \partial_t r^{11}_i(\epsilon) =-2iA \frac{(e v_{Fi})^2}{3}\int d\epsilon' [l_i(\epsilon, \epsilon')^2 \operatorname{Im} F^{11}_i(\epsilon, \epsilon') - p_i(\epsilon,\epsilon')^2 \operatorname{Im} F^{21}_i(\epsilon, \epsilon')]W_i(\epsilon, \epsilon')
	\label{eqn:r11}\,,
\end{equation}
\begin{equation}
	r^{22}_i(\epsilon) = -  r^{11}_i(\epsilon)
	\label{eqn:r22}
\end{equation}
The terms $r^{11}_i$ and $r^{22}_i$ correspond to the quasiparticle excitations. The remaining $r^{ab}_i$ terms correspond to the collective modes. We break up the remaining terms into odd and even components, which are responsible for the Higgs and Leggett modes respectively.
\begin{equation}
	r^{21}_i(\epsilon) = r^{21,\text{odd}}_{i}(\epsilon)  + r^{21,\text{even}}_{i}(\epsilon) \,,
\end{equation}
\begin{align}
	r^{21,\text{odd}}_{i}(-\epsilon)  &= - r^{21,\text{odd}}_{i}(\epsilon) \,,\\
	r^{21, \text{even}}_{i}(\epsilon)  &= r^{21,\text{even}}_{i}(-\epsilon) \,.
\end{align}
The equations of motion for these terms are
\begin{align}
	\left[i \partial_t - 2E \right]r^{21,\text{odd}}_{i}(\epsilon) &= -(1-2f)(u_i(\epsilon)^2-v_i(\epsilon)^2)\delta \Delta_i'\bigg|_2
	\notag \\
	&\qquad -2A\frac{(ev_{Fi})^2}{3}\int d\epsilon' W_i(\epsilon, \epsilon')l_i(\epsilon,\epsilon')p_i(\epsilon, \epsilon')[F_i^{21}(\epsilon,\epsilon') - F_i^{22}(\epsilon,\epsilon')]		\label{eqn:r21odd}\\
	\left[i \partial_t - 2E \right]r^{21,\text{even}}_{i}(\epsilon) &= (1-2f)\frac{\Delta_i^{\text{eq}} s_i e^2 \textbf{A}^2}{2m_iE} - i(1-2f)\delta \Delta_i''\bigg|_2
	\label{eqn:r21even}.
\end{align}
After exchanging momentum sums into energy integrals, the gap equations written in terms of the angle averaged quasiparticles become
\begin{align}
	\delta \Delta_i'\big|_2 &= \sum_j U_{ij}N_j(0) \int d\epsilon \left\lbrace -u_j(\epsilon)v_j(\epsilon) \left(r^{11}_j(\epsilon) - r^{22}_j(\epsilon) \right) + \frac{1}{2}\left(u_j(\epsilon)^2 - v_j(\epsilon)^2 \right) \left(r^{21}_j(\epsilon) + r^{12}_j(\epsilon) \right)\right\rbrace \label{eqn:gap2'}\,,\\
	\delta \Delta_i''\big|_2 &= \sum_j U_{ij}N_j(0) \int d\epsilon \frac{1}{2i}\left(r^{21}_j(\epsilon) - r^{12}_j(\epsilon) \right) \label{eqn:gap2''} .
\end{align}
The equations of motion are solved numerically, and must be solved self-consistently with the gap equations at each time step. This condition induces the collective modes.

Finally, to consider pump-probe simulations and THG, we must go to third order. First, the diamagnetic third order current can be directly calculated from the angle-averaged quantities $r_i^{ab}$.
 \begin{eqnarray}
	\mathbf{j}_D &=& -
	\sum_{i}^{}
	\int_{}^{}d\epsilon
	\frac{s_i e^2 \mathbf{A}}{m_i}
	\left[
		\left( u_i(\epsilon)^2 - v_i(\epsilon)^2 \right)
		\left(
			r^{11}_i(\epsilon)
			-
			r^{22}_i(\epsilon)
		\right)
		+ 
		2 u(\epsilon)_iv(\epsilon)_i
		\left(
			r^{21}_i(\epsilon) 
			+
			r^{12}_i(\epsilon)
		\right)
	\right]\,.
 \end{eqnarray}
To find the paramagnetic third order current, we start from the third order equation of motion
\begin{align}
i \partial_t \rho_{i\mathbf{k}\mathbf{k}'}\big|_3
  &=
  \left( 
    H_{i\mathbf{k'k'}}^{(1)} - H_{i\mathbf{kk} }^{(2)} 
  \right)\bigg|_0
    \rho_{i\mathbf{kk'}}\big|_3
    +
  \sum_{\mathbf{q}}^{}
  \left( 
    H_{i\mathbf{k'q}}^{(1)}\big|_1 \rho_{i\mathbf{kq }}\big|_2 -
    H_{i\mathbf{qk} }^{(2)}\big|_1 \rho_{i\mathbf{qk'}}\big|_2
  \right)
  +
  \left( 
    H_{i\mathbf{k'k'}}^{(1)} - H_{i\mathbf{kk} }^{(2)} 
  \right)\bigg|_2
    \rho_{i\mathbf{kk'}}\big|_1
.\end{align}
We proceed by computing the equation explicitly for $\rho^{11}_{i \mathbf{kk'}}$. Here, we do not consider any contributions from the Leggett mode (i.e. terms involving $\delta\Delta_i''$) or the $\mathbf{A}^2$ part of the EM field. These terms vanish because of particle hole symmetry. The remaining contributions, consisting of quasiparticles and Higgs mode, are

\begin{eqnarray}
	\left[i\partial_t - \left( E_{i\mathbf{k'}}-E_{i\mathbf{k}} \right)\right]
	\rho_{i\mathbf{kk}'}^{11} \big\rvert_3 
	&=&
	\mathbf{J}_{i\mathbf{k'k}} \cdot \mathbf{e} \,
	l_{i\mathbf{k'k}} \left\{
		A \left[
			\rho_{i\mathbf{k'k'}}^{11}\big\rvert_2
			-\rho_{i\mathbf{kk}}^{11}\big\rvert_2
		\right]
		+
		\delta\Delta_i'\big\rvert_2
		\left(\frac{\Delta_i}{E_{i\mathbf{k'}}}-\frac{\Delta_i}{E_{i\mathbf{k}}}\right) 
		F_{i\mathbf{kk'}}^{11}
	\right\}
	\notag\\
	&&+
	\mathbf{J}_{i\mathbf{k'k}} \cdot \mathbf{e} \,
	p_{i\mathbf{k'k}} \left\{
		A
		\left[
			\rho_{i\mathbf{k'k'}}^{21}\big\rvert_2
			+\rho_{i\mathbf{kk}}^{12}\big\rvert_2
		\right]
		+
		\delta\Delta_i'\big\rvert_2
		\left(
			-\frac{\epsilon_{i\mathbf{k'}}}{E_{i\mathbf{k'}}}
			F_{i\mathbf{kk'}}^{12}
			-\frac{\epsilon_{i\mathbf{k}}}{E_{i\mathbf{k}}}
			F_{i\mathbf{kk'}}^{21}	
		\right) 
	\right\}\,.
\end{eqnarray}
The next step is to insert this into the expression for the paramagnetic current (\ref{eqn:jp}). Replacing summation over $\mathbf{k}$ with integrals over energy, we find
\begin{eqnarray}
	\mathbf{j}_P\big\rvert_3 &=& \mathbf{e} \sum_{i\mathbf{k}\mathbf{k}'}^{}
	\mathbf{e}\cdot \mathbf{J}_{i\mathbf{kk'}}\left[l_{i\mathbf{kk'}}
	\left(
		\rho_{i\mathbf{kk'}}^{11}\big\rvert_3
		+ 
		\rho_{i\mathbf{kk'}}^{22}\big\rvert_3
	\right)
	+ p_{i\mathbf{kk'}}
	\left(
		\rho_{i\mathbf{kk'}}^{21}\big\rvert_3
		-
		\rho_{i\mathbf{kk'}}^{12}\big\rvert_3
	\right)
	\right]
	\notag\\
	&=& 
	\mathbf{e} \sum_{i}^{} N_i(0) \int d\epsilon d\epsilon'
	\left[l_{i}(\epsilon,\epsilon')
	\left(
		\langle \mathbf{e}\cdot\mathbf{J}_{i\mathbf{kk'}}\rho_{i\mathbf{kk'}}^{11}\rangle_{\text{Av}}\big\rvert_3
		+ 
		\langle\mathbf{e}\cdot\mathbf{J}_{i\mathbf{kk'}}\rho_{i\mathbf{kk'}}^{22}\rangle_{\text{Av}}\big\rvert_3
	\right) \right. \notag \\
	&+& 
	\left.
	p_i(\epsilon,\epsilon')
	\left(
		\langle\mathbf{e}\cdot\mathbf{J}_{i\mathbf{kk'}}\rho_{i\mathbf{kk'}}^{21}\rangle_{\text{Av}}\big\rvert_3
		-
		\langle\mathbf{e}\cdot\mathbf{J}_{i\mathbf{kk'}}\rho_{i\mathbf{kk'}}^{12}\rangle_{\text{Av}}\big\rvert_3
	\right)
	\right]
\end{eqnarray}
We see that in fact only the angle averaged quantities
\begin{eqnarray}
	\langle \mathbf{e}\cdot\mathbf{J}_{i\mathbf{kk'}}\rho_{i\mathbf{kk'}}^{ab}\rangle_{\text{Av}} = 
	\int 
	\frac{d \Omega_{\mathbf{k}}}{4\pi}
	\frac{d \Omega_{\mathbf{k}'}}{4\pi }
	\mathbf{e}\cdot\mathbf{J}_{i\mathbf{kk'}}\rho_{i\mathbf{kk'}}^{ab}
\end{eqnarray}
occur. The differential equation for these quantities is 
\begin{align}
	\left[i\partial_t - \left( E'-E \right)\right]
	\langle 
	\mathbf{e} \cdot \mathbf{J}_{i\mathbf{kk'}}	
	\rho_{i\mathbf{kk'}}^{11} 
	\rangle_{\text{Av}}
	\big\rvert_3 
	&=
	\langle
	\left|
	\mathbf{J}_{i\mathbf{k'k}} \cdot \mathbf{e}
	\right|^2
	\,
	l_{i\mathbf{k'k}} \left\{
		A \left[
			\rho_{i\mathbf{k'k'}}^{11}\big\rvert_2
			-\rho_{i\mathbf{kk}}^{11}\big\rvert_2
		\right]
		+
		\delta\Delta_i'\big\rvert_2
		\left(\frac{\Delta_i}{E_{i\mathbf{k'}}}-\frac{\Delta_i}{E_{i\mathbf{k}}}\right) 
		F_{i\mathbf{kk'}}^{11}
	\right\}
	\rangle_{\text{Av}}
	\nonumber
	\\
	&+
	\langle
	\left|
	\mathbf{J}_{i\mathbf{k'k}} \cdot \mathbf{e}
	\right|^2
	\,
	p_{i\mathbf{k'k}} \left\{
		A
		\left[
			\rho_{i\mathbf{k'k'}}^{21}\big\rvert_2
			+\rho_{i\mathbf{kk}}^{12}\big\rvert_2
		\right]
		+
		\delta\Delta_i'\big\rvert_2
		\left(
			-\frac{\epsilon_{i\mathbf{k'}}}{E_{i\mathbf{k'}}}
			F_{i\mathbf{kk'}}^{12}
			-\frac{\epsilon_{i\mathbf{k}}}{E_{i\mathbf{k}}}
			F_{i\mathbf{kk'}}^{21}	
		\right) 
	\right\}
	\rangle_{\text{Av}}
\end{align}
Now we make the final approximation
\begin{eqnarray}
	\langle
	\left|
	\mathbf{J}_{i\mathbf{k'k}} \cdot \mathbf{e}
	\right|^2
	\rho_{i\mathbf{kk'}}^{ab} \big\rvert_2
	\rangle_{\text{Av}}
	\approx
	\langle
	\left|
	\mathbf{J}_{i\mathbf{k'k}} \cdot \mathbf{e}
	\right|^2
	\rangle_{\text{Av}}
	\langle
	\rho_{i\mathbf{kk'}}^{ab} 
	\rangle_{\text{Av}}\big\rvert_2\,.
\end{eqnarray}
With this, the differential equation becomes
\begin{align}
	\left[i \partial_t - \left( E'-E \right)\right]
	\frac{\langle 
	\mathbf{e} \cdot \mathbf{J}_{i\mathbf{kk'}}	
	\rho_{i\mathbf{kk'}}^{11} 
	\rangle_{\text{Av}}
	\big\rvert_3}
	{\langle
	\left|
	\mathbf{J}_{i\mathbf{kk'}} \cdot \mathbf{e}
	\right|^2
	\rangle_{\text{Av}}}
	&=
	l_i(\epsilon',\epsilon) \left\{
		A \left[
			\langle\rho_{i\mathbf{k'k'}}^{11}\rangle_{\text{Av}}\big\rvert_2
			-\langle\rho_{i\mathbf{kk}}^{11}\rangle_{\text{Av}}\big\rvert_2
		\right]
		+
		\delta\Delta_i'\big\rvert_2
		\left(\frac{\Delta_i}{E'}-\frac{\Delta_i}{E}\right) 
		F_i(\epsilon,\epsilon')^{11}
	\right\} \nonumber
	\\
	+&
	p_i(\epsilon',\epsilon) \left\{
		A
		\left[
			\langle\rho_{i\mathbf{k'k'}}^{21}\rangle_{\text{Av}}\big\rvert_2
			+\langle\rho_{i\mathbf{kk}}^{12}\rangle_{\text{Av}}\big\rvert_2
		\right]
		+
		\delta\Delta_i'\big\rvert_2
		\left(
			-\frac{\epsilon'}{E'}
			F_i(\epsilon,\epsilon')^{12}
			-\frac{\epsilon}{E}
			F_i(\epsilon,\epsilon')^{21}	
		\right) 
	\right\}\,.
\end{align}
Note that $\langle\left|\mathbf{J}_{i\mathbf{k'k}} \cdot \mathbf{e} \right|^2 \rangle_{\text{Av}}=\langle\left|\mathbf{J}_{i\mathbf{kk'}}
\cdot \mathbf{e} \right|^2 \rangle_{\text{Av}}$. Defining 
\begin{eqnarray}
	R^{ab}_i(\epsilon,\epsilon') &=& 
	\frac{1}{\langle \left| \mathbf{J}_{i\mathbf{kk'}} \cdot \mathbf{e} \right|^2 \rangle_{\text{Av}}}
	\langle 
	\mathbf{e} \cdot \mathbf{J}_{i\mathbf{kk'}}	
	\rho_{i\mathbf{kk'}}^{ab} 
	\rangle_{\text{Av}}
	\big\rvert_3 ,
\end{eqnarray}
and noting
\begin{eqnarray}
	r^{ab}_i(\epsilon) &=& 
	\langle\rho_{i\mathbf{kk}}^{ab}\rangle_{\text{Av}}\big\rvert_2 ,
\end{eqnarray}
we rewrite this as
\begin{eqnarray}
	\left[i\partial_t - \left( E'-E \right)\right]
	R_i^{11}(\epsilon,\epsilon')
	\big\rvert_3 
	&=&
	l_i(\epsilon',\epsilon)\left\{
		A \left[
			r_i^{11}(\epsilon')
			-
			r_i^{11}(\epsilon)
		\right]
		+
		\delta\Delta_i'\big\rvert_2
		\left(\frac{\Delta_i}{E'}-\frac{\Delta_i}{E}\right) 
		F_i^{11}(\epsilon,\epsilon')
	\right\}
	\notag\\
	&&+
	p_i(\epsilon',\epsilon) \left\{
		A
		\left[
			r_i^{21}(\epsilon')
			+
			r_i^{12}(\epsilon)
		\right]
		+
		\delta\Delta_i'\big\rvert_2
		\left(
			-\frac{\epsilon'}{E'}
			F_i^{12}(\epsilon,\epsilon')
			-\frac{\epsilon}{E}
			F_i^{21}(\epsilon,\epsilon')	
		\right) 
	\right\}
\end{eqnarray}
which precisely gives Eq.~48 in \cite{Murotani2019}. Similar equations can be derived for the remaining $R_i^{ab}$.

Rewriting the expression for the paramagnetic current $\mathbf{j}_P$ in terms of the $R_i^{ab}(\epsilon,\epsilon')$, and then making the Mattis-Bardeen impurity replacement (\ref{eq:MB}), yields

\begin{eqnarray}
	\mathbf{j}_P\big\rvert_3 
	&=& 
	\mathbf{e} \sum_i
	\frac{e^2 n_i}{2m_i}
	\int d\epsilon d\epsilon'
	W_i(\epsilon,\epsilon')
	\left[l_i(\epsilon,\epsilon')
	\left(
		R_i^{11}(\epsilon,\epsilon')
		+ 
		R_i^{22}(\epsilon,\epsilon')
	\right)
	+ p_i(\epsilon,\epsilon')
	\left(
		R_i^{21}(\epsilon,\epsilon')
		-
		R_i^{22}(\epsilon,\epsilon')
	\right)
	\right]\,.
\end{eqnarray}

\end{widetext}

\bibliography{literature}

\end{document}